\documentclass{article}

\usepackage{booktabs} 
\usepackage{tabularx} 
\usepackage{hyperref}
\usepackage{algorithm}
\usepackage{subcaption} 
\usepackage{xcolor,pifont}
\usepackage{tikz}
\usepackage{ulem}
\usepackage{url}
\usepackage{soul}
\usepackage[most]{tcolorbox}
\usepackage{arxiv}
\usepackage{svg}
\usepackage{pgf}
\usepackage[utf8]{inputenc} 
\usepackage[T1]{fontenc}    
\usepackage{hyperref}       
\usepackage{url}            
\usepackage{booktabs}       
\usepackage{amsfonts}       
\usepackage{nicefrac}       
\usepackage{microtype}      
\usepackage{lipsum}		
\usepackage{graphicx}
\usepackage[square,sort,comma,numbers]{natbib}

\usepackage{doi}
\definecolor{green}{rgb}{0.0, 0.5, 0.0} 
\newcommand*\colourcheck[1]{%
  \expandafter\newcommand\csname #1check\endcsname{\textcolor{#1}{\ding{52}}}%
}

\newcommand*\colourcross[1]{%
  \expandafter\newcommand\csname #1cross\endcsname{\textcolor{#1}{\ding{56}}}%
} 
\colourcheck{green}
\colourcross{red}

\usepackage{colortbl}
\usepackage{booktabs}
\newcommand{\boxhl}[2]{%
    {\sethlcolor{#1!20}\hl{#2}}%
}

\def\BibTeX{{\rm B\kern-.05em{\sc i\kern-.025em b}\kern-.08em
    T\kern-.1667em\lower.7ex\hbox{E}\kern-.125emX}}

\title{Driving Towards Inclusion: A Systematic Review
of AI-powered Accessibility Enhancements for
People with Disability in Autonomous Vehicles}

\author{
  Ashish Bastola \\
  Clemson University \\
  Clemson, SC, USA\\
  \texttt{abastol@clemson.edu} \\
  \And
  Hao Wang \\
  Clemson University \\
  Clemson, SC, USA\\
  \texttt{hao9@clemson.edu} \\
    \And
  Sayed Pedram Haeri Boroujeni \\
  Clemson University \\
  Clemson, SC, USA\\
  \texttt{shaerib@clemson.edu} \\
    \And
  Julian Brinkley \\
  Clemson University \\
  Clemson, SC, USA\\
  \texttt{jbrinkl@clemson.edu} \\
    \And
  Ata Jahangir Moshayedi\\
  Jiangxi University of Science and Technology \\
  Ganzhou, China\\
  \texttt{ajm@jxust.edu.cn} \\
    \And
  Abolfazl Razi \\
    Clemson University\\
   Clemson, SC, USA\\
  \texttt{arazi@clemson.edu} \\
}

\renewcommand{\shorttitle}{Driving Towards Inclusion: A Systematic Review of AI-powered Accessibility Enhancements . }

\hypersetup{
pdftitle={Driving Towards Inclusion: A Systematic Review
of AI-powered Accessibility Enhancements for
People with Disability in Autonomous Vehicles},
pdfauthor={Ashish Bastola},
pdfkeywords={Autonomous Vehicles, In-vehicle Systems, AI-based Accessibility Accommodation, Human-Computer
Interaction},
}

\begin{document}
\maketitle

\begin{abstract}

This paper provides a comprehensive and, to our knowledge, the first review of inclusive human-computer interaction (HCI) within autonomous vehicles (AVs) and human-driven cars with partial autonomy, emphasizing accessibility and user-centered design principles.  
We explore the current technologies and HCI systems designed to enhance passenger experience, particularly for individuals with accessibility needs. Key technologies discussed include brain-computer interfaces, anthropomorphic interaction, virtual reality, augmented reality, mode adaptation, voice-activated interfaces, haptic feedback, etc. Each technology is evaluated for its role in creating an inclusive in-vehicle environment. Furthermore, we highlight recent interface designs by leading companies and review emerging concepts and prototypes under development or testing, which show significant potential to address diverse accessibility requirements. Safety considerations, ethical concerns, and adoption of AVs are other major issues that require thorough investigation. Building on these findings, we propose an end-to-end design framework that addresses accessibility requirements across diverse user demographics, including older adults and individuals with physical or cognitive impairments.
This work provides actionable insights for designers, researchers, and policymakers aiming to create safer and more comfortable environments in autonomous and regular vehicles accessible to all users. 

\end{abstract}


\label{sec:introduction}

The evolution of Autonomous Vehicles (AVs) has opened unprecedented opportunities to reshape the future of transportation. Beyond their potential to improve efficiency and safety, AVs provide a significant opportunity for enhancing mobility and independence for individuals with disabilities. Central to this vision is the integration of inclusive Human-Computer Interaction (HCI) systems, which ensure accessibility and usability needs are fulfilled for all passengers, regardless of their physical or cognitive abilities. Similarly, most human-driven vehicles fail to cover the diverse needs of drivers and passengers with partial physical and cognitive impairments, thereby undermining their \textit{inclusivity}. 
Despite advancements in Artificial Intelligence (AI) and user interface technologies, accessibility remains an underexplored aspect of both regular vehicle and AV system design. This underscores the need for HCI systems that provide fair,  convenient, and inclusive accessibility for individuals with varying levels of ability across diverse demographics.

\subsection{Motivation} 

\begin{figure}
    \centering
    \includegraphics[width=\linewidth]{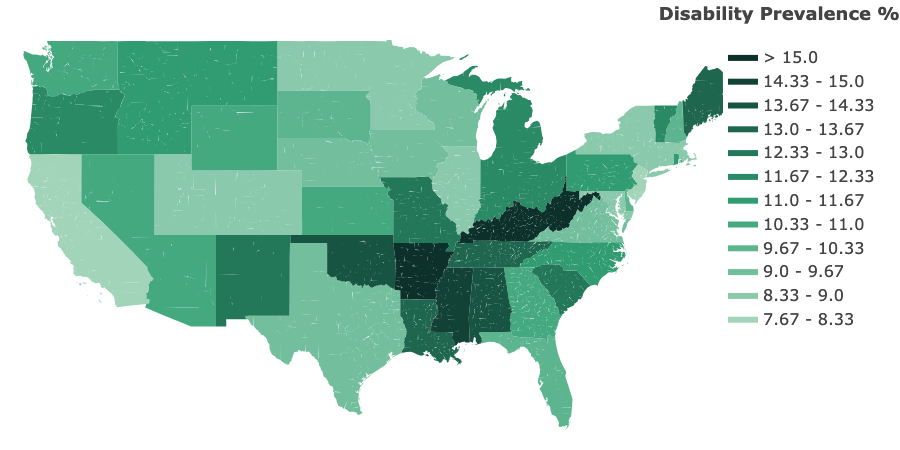}
    \caption{Disability Prevalence (\%) in United States}
    \label{fig:disability-prevalance}
\end{figure}

\begin{figure*}[ht!]
    \centering
    \includegraphics[width=0.9\linewidth]{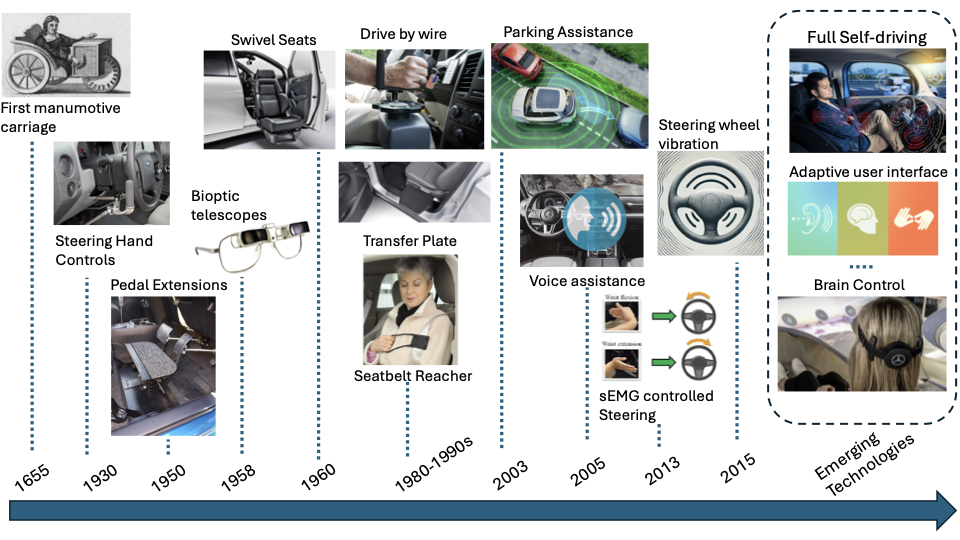}
    \caption{Timeline of various in-vehicle accessible technologies}
    \label{fig:timeline}
\end{figure*}

According to World Health Organization (WHO) statistics, more than a billion people worldwide have some form of disability \cite{disablePopulation}. Around 16.6 percent of the population of the entire United States aged above 65, thus requiring special accommodations \cite{olderPop}. Mobility-related issues are common among these groups, which significantly affects their quality of life \cite{rosso2013mobility, sze2017access, cordts2021mobility, alriksson2007quality, metz2000mobility}. Every state in the US has high disability prevalence, with the minimum being 7.67\% and the maximum above 15\% averaging around 10\% based on the 2018 5-year estimates from the American Community survey\cite{USCensusBureau2022} as shown in Figure \ref{fig:disability-prevalance}. Among all individuals who never leave their homes, 54 percent are the ones with disabilities, and about half a million of those individuals indicate they never leave their homes because of transportation difficulties \cite{leavingHome}. Fortunately, AVs are becoming increasingly more common and are expected to revolutionize the transportation industry in the upcoming years\cite{fagnant2015preparing, rajasekhar2015autonomous, anderson2014autonomous}. AVs can potentially enhance these individuals' lives substantially \cite{guerreiro2019cabot, kassens2021autonomous, bennett2019willingness, bennett2020willingness}. A timeline of various in-vehicle accessibility technologies dating back from their early days to today's contemporary systems is shown in Figure \ref{fig:timeline}.

\subsection {Accessibility Challenges and Legislative Gaps} 

Although this technological leap has induced significant hope among individuals unable to use current transportation systems, the design decision is crucial to ensure these technologies remain accessible in the long run with evolving needs, standards, and societal conditions.

\cite{brinkley2017opinions, choi2015investigating, vosooghi2019shared, ackermann2019experimental}. Moreover, several studies show a small percentage of research on human-machine interface (HMI) 
research targets under-explored user groups such as the visually impaired and older adults, highlighting the need for broader accessibility in-vehicle technology to improve safety and independence\cite{dong2024review}.

Our study emphasizes the need to re-evaluate in-vehicle systems and components, advocating for strategies that enhance their \textit{inclusivity} for these user groups. We also highlight the absence of comprehensive federal laws in the US governing AVs for these groups. Over the years, there has been a gradual increase in the number of states considering legislation related to AVs, but such legislation has not adequately addressed the needs of people with disabilities \cite{legislation}. This poses a significant difficulty among these individuals when AVs are deployed as Mobility as a Service (MaaS) \cite{bagloee2016autonomous}.

\subsection{Approaches} 

Standardizing these technologies is crucial to enable these groups of individuals to access and use these services \cite{hancock2020challenges}. There exists some recent literature focusing on the AVs' applicability to enhance the quality of life; however, there is a lack of emphasis on how these technologies can be made inclusive \cite{politis2018designing, politis2021designing, nanchen2022perceptions}. We conduct this review to bring this into perspective and identify current technological gaps that make these systems inaccessible. We also incorporated methods that are not already compliant with inclusive design but have the potential to be adapted and scaled to facilitate improved interaction mechanisms for people with disability. 

We also investigate barriers to making existing technology more inclusive. In our assessments, We account for the end user's interactions with the vehicle, from ingress to egress, considering all the challenges an individual might face and potential solutions.

\begin{figure}[ht!]
   \centering
   \includegraphics[width=0.55\linewidth, height=0.9\textheight]{images/organization_of_paper.jpg}
   \caption{The organization of this review paper.}
   \label{fig:organization}
\end{figure}

\subsection{Review Criteria}

This review aims to explore diverse user needs through a careful analysis of influential literature that has significantly shaped the field.
Following the PRISMA's systematic review protocol \cite{page2021prisma}, we conducted a multi-stage literature screening. Our initial search across Google Scholar, IEEE, and ACM databases yielded over 1,300 records. After applying exclusion criteria focused on relevance, language, accessibility, and human-centered design, we filtered the results to over 100  highly relevant publications. These were subjected to a thematic analysis, following Braun and Clarke’s approach \cite{braun2006using}, where publications were coded iteratively to identify emerging themes. The resulting dataset, primarily sourced from IEEE, ACM, and Elsevier, forms the foundation of this review, ensuring a comprehensive analysis of accessibility and inclusion for in-vehicle HCI for AVs. In addition to these, we also considered several technical reports, government reports, road maps, statistics, and recent feature announcements from company websites and updates relevant to the review.
This curated review thus highlights gaps in existing literature and suggests directions for future research, providing a foundational resource for developing accessible AV systems.

\subsection{Existing Reviews} 

Most reviews that explore AV technologies often overlook accessibility and HCI requirements, merely focusing on the technical aspects. On the other hand, HCI reviews that consider accessibility and inclusion lack vehicular context, disregarding subtle details on AV features. Our review is the first to bridge this gap by bringing both aspects into perspective while summarizing the recent technological innovations. For our review, we adhere to the standard framework that defines all vehicles as \textbf{'Autonomous'}, with varying levels of autonomy as defined by the Society of Automotive Engineers (SAE)\cite{sae_j3016_update}. Under this definition, Level 0 refers to vehicles with no automation, while Level 5 represents full automation, thus considering regular vehicles with varying levels of partial autonomy. 

In this section, we summarize the key highlights of each related review paper to assess whether or not specific topics have been thoroughly covered. Table \ref{table:surveys} highlights the major contents covered by each review paper and its potential limitations and technical gaps. 
IVI in this table, denotes the In-vehicle Interactions. The focus of this table is on highly cited and most relevant papers. The publication year of these articles ranged from 2017 to current. To our surprise, many review papers lacked sufficient details on some core concepts of in-vehicle interactions, such as Ethics, Trust, Safety, User Acceptance, adoption of autonomous driving in society, and features implemented via Vehicle-to-Everything (V2X) interactions. 

Furthermore, we noticed that almost all of these review papers missed \textit{personalization}, one of the most crucial factors regarding disability. 
Disability types are inherently different regarding user interaction; hence ignoring the personalization factor generates a bias in these technologies being favorable to specific disability types and rendering them useless for others. Almost all of this literature discussed user experience without this consideration, which greatly misses out on requirements and personalization needs for different disability types. Another important aspect that was greatly missing in this literature was vehicle-to-everything interaction. We noticed literature that discussed in-vehicle interaction greatly missed out on interaction with external communication, such as with pedestrians and other road objects or vehicles. Situational awareness is a major aspect of in-vehicle interaction, and safety doesn't refer to those inside but also to others outside. In-vehicle users should be able to get enough information about the surroundings to ensure a comfortable user experience. In case of an unforeseen emergency, in-vehicle users should be able to find the best way to ensure safety. When considering vulnerable user groups, ensuring every vehicle action to provide relief is important. Unfavorable vehicle actions without clear information to the in-vehicle user can easily lead to feelings of shock and anxiety. Also, they could be at risk of post-traumatic stress disorder(PTSD) even though psychological impacts can vary greatly depending on the circumstances and the individual's resilience. At minimum user's experience is greatly affected failing to consider these factors. 

\begin{figure}[ht]
    \centering
    \includegraphics[width=1\linewidth]{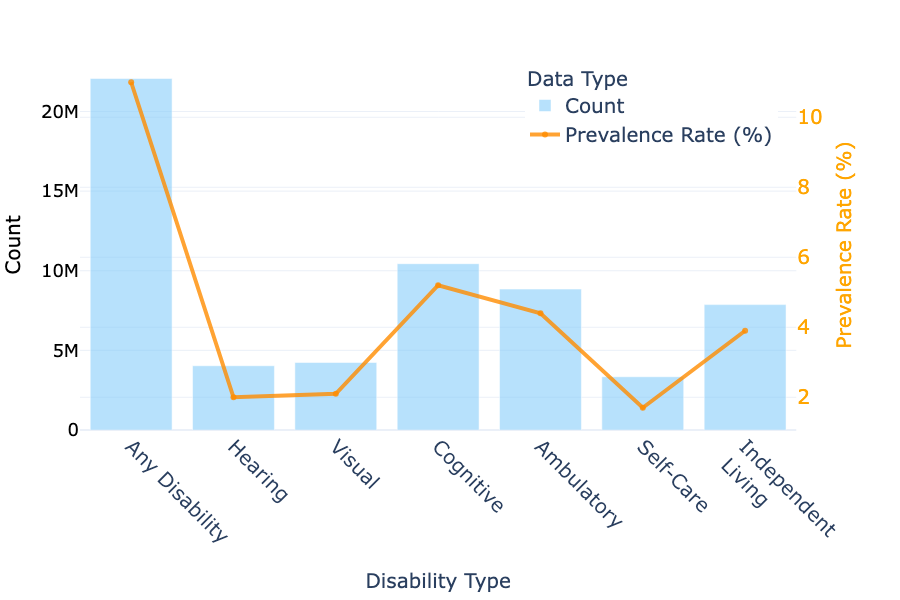}
    \caption{Disability Types: Count and Prevalence Rate for 2022 for working age 18-64 based on \cite{ACS2022PUMS}. }
    \label{fig:enter-label}
\end{figure}

\begin{table*}[!htbp]
\fontsize{8.2}{8.8}\selectfont

\caption{Summary of HMI design.}
\label{table:surveys}
\begin{tabularx}{\textwidth}{p{0.02\textwidth}p{0.06\textwidth}p{0.31\textwidth}p{0.32\textwidth}p{0.315\textwidth}}

\hline
\textbf{Year} & \textbf{Surveys} & \textbf{Content} & \textbf{Gaps} & \textbf{Highlights} \\
\hline
\addlinespace
2024 & \cite{dong2024review} & 
        Classification of stages of HMI design process \newline
        HMI Design trends and strategies \newline
        HMI considerations for under-explored users and specialized vehicles.
    & 
        Recent Autonomy, AI features and their effectiveness \newline
        Role of user interaction in trust and safety \newline
        Analysis of disability needs and their solutions \newline & 
        \greencheck IVI \redcross Ethics \redcross V2X \redcross Trust\newline
        \redcross Acceptance \redcross Safety \redcross AI\newline
        \redcross Disability needs \redcross Autonomy \newline
        \redcross Mode Adaptation \greencheck User Exp. \\

2024 & \cite{mandujano2024human} & 
        Historical and current development of HMI \newline
        Emerging trends and technological developments in HMIs for AVs \newline
        Challenges of HMI research strategies
        & 
        Analysis of disability needs for specific users\newline
        Disability needs and mode adaptation \newline
        HMI Design Principles, Multi-modal user Interaction, Ethical Implications \newline
        & 
        \greencheck IVI \redcross Ethics \greencheck V2X \redcross Trust\newline
        \redcross Acceptance \greencheck Safety \greencheck AI\newline
        \redcross Disability needs \greencheck Autonomy\newline
        \redcross Mode Adaptation \greencheck User Exp. \\

2023 & \cite{yan2023user}& 
        Analysis of the influence of HMI on user acceptance\newline
        Impact of External HMI on VRU\newline
        Control transfer between vehicle and user.
        & 
        Analysis of disability needs for diverse users\newline
        Limited focus on regulatory and ethical issues\newline
        Mode-confusion, driver adaptation, and driver readiness during NDRA 
        & 
        \greencheck IVI \greencheck Ethics \redcross V2X \greencheck Trust\newline
        \greencheck Acceptance \greencheck Safety \redcross AI\newline
        \redcross Disability needs \greencheck Autonomy\newline
        \redcross Mode Adaptation \greencheck User Exp.\\
\addlinespace

2023 & \cite{mourtzis2023future} & 
    The emphasis of Industry 5.0 is blending Human and AI capabilities\newline
    Human \& machine capabilities to enhance HMI\newline
    The potential of digital twins to enhance UI
    & 
    Insufficient focus in AV applications\newline
    No mention of accessibility and disability needs\newline
    Insufficient focus on ethical implications and Trust
    & 
    \redcross IVI \redcross Ethics \redcross V2X \redcross Trust\newline
    \redcross Acceptance \greencheck Safety \greencheck AI\newline
    \redcross Disability needs \redcross Autonomy\newline
    \redcross Mode Adaptation \greencheck User Exp.\\
\addlinespace
    
2022 & \cite{stampf2022towards} & 
    Implicit inputs (physiological, kinesthetic, auditory) to infer user states\newline
    Impact of automation on implicit interaction and state recognition\newline
    Research gaps in implicit interaction for AV& 
    
    Most state recognition methods target manual driving scenarios\newline
    Limited focus on implicit input's role in UX and trust\newline
    Lacks framework to minimize user state errors.
    & 
    \greencheck IVI \redcross Ethics \redcross V2X \greencheck Trust\newline
    \redcross Acceptance \greencheck Safety \redcross AI\newline
    \redcross Disability needs \greencheck Autonomy\newline
    \redcross Mode Adaptation \greencheck User Exp. \\
\addlinespace
    
    2022 & \cite{lee2022systematic} & 
    Analyze objectives and design of in-vehicle agents \newline
    Effect of in-vehicle agent on driver\newline
    Present design guidelines for effective in-vehicle interaction
    & 
    No mention of accessibility and disability needs\newline
    Omits ethical implications of in-vehicle agents\newline
    Insufficient exploration of NDRA, trust, and acceptance
    & 
    \greencheck IVI \redcross Ethics \redcross V2X \greencheck Trust\newline
    \redcross Acceptance \greencheck Safety \greencheck AI\newline
    \redcross Disability needs \greencheck Autonomy\newline
    \redcross Mode Adaptation \greencheck User Exp. \\
\addlinespace

2021 & \cite{detjen2021increase} & 
    Progress in automotive UI and its potential to facilitate higher automation\newline
    Designing with user needs for AV acceptance \newline
    Explore design space for future IVI.
    & 
    No account for accessibility and disability needs\newline
    Minimal mention of ethical implications\newline
    There is no mention of the mode adaptation
    & 
    \greencheck IVI \redcross Ethics \redcross V2X \greencheck Trust\newline
    \greencheck Acceptance \greencheck Safety \greencheck AI\newline
    \redcross Disability needs \greencheck Autonomy\newline
    \redcross Mode Adaptation \greencheck User Exp.\\

\addlinespace

2021 & \cite{dicianno2021systematic} & 
     AVs' potential for accessible travel\newline
    Importance of inclusive design for AVs\newline
    Importance of educating stakeholders about AV accessibility issues
    & 
    Minimal focus on ethics and AV acceptance\newline
    There is no mention of mode adaptation based on disability needs\newline
    No mention of engagement in NDRA
    & 
    \greencheck IVI \redcross Ethics \redcross V2X \greencheck Trust\newline
    \redcross Acceptance \greencheck Safety \redcross AI\newline
    \greencheck Disability needs \greencheck Autonomy\newline
    \redcross Mode Adaptation \greencheck User Exp.\\

\addlinespace

2021 & \cite{fink2021fully} & 
    Need for AVs to be accessible to BVIs \newline
    Outlines policy and legislative recommendations to ensure AVs are inclusive\newline
    Linking HMI with a smartphone for accessibility
    & 
    Guidelines for designing interfaces for BVI users\newline
    Strategy for integration within legal frameworks\newline
    It focuses solely on visual disabilities and does not address other disability groups.
    & 
    \greencheck IVI \redcross Ethics \redcross V2X \greencheck Trust\newline
    \greencheck Acceptance \greencheck Safety \greencheck AI\newline
    \greencheck Disability needs \greencheck Autonomy\newline
    \redcross Mode Adaptation \greencheck User Exp.\\

\addlinespace
2021 & \cite{riegler2021systematic} & 
    Application areas of VR for HMI research\newline
    Highlights VR use in evaluating HMIs\newline
    Recommendation for VR study design in driving automation research
    & 
    There is minimal mention of the ethical and social implications of automated driving in VR\newline
    Focuses only on VR, not other interactions\newline
    No mention of VR accessibility for disabled\newline
    & 
    \greencheck IVI \greencheck Ethics \redcross V2X \greencheck Trust\newline
    \greencheck Acceptance \greencheck Safety \redcross AI\newline
    \redcross Disability needs \greencheck Autonomy\newline
    \redcross Mode Adaptation \greencheck User Exp.\\

\addlinespace

2020 & \cite{marcano2020review} & 
    Shared control in lane keeping, obstacle avoidance, and control transitions.\newline
    Analyze steering control algorithm design\newline
    Focuses on steering conflicts and adaptation\newline
    & 
    No mention of accessibility and disability needs\newline
    No mention of regulatory and ethical implications\newline
    Minimal mention of trust, situational awareness
    & 
    \greencheck IVI \greencheck Ethics \redcross V2X \greencheck Trust\newline
    \greencheck Acceptance \greencheck Safety \redcross AI\newline
    \redcross Disability needs \greencheck Autonomy\newline
    \redcross Mode Adaptation \greencheck User Exp.\\

\addlinespace

2017 & \cite{young2017toward} & 
    Assess IVIS, ADAS design for age groups\newline
    Role of system design to support older drivers \newline
    Considers decline in physical, sensory, and cognitive functions in HMI design
    & 
    Lacks detailed review of hearing impairment\newline
    There is no mention of ethics or AV acceptance\newline
    No mention of NDRA, trust, and situational awareness
    & 
    \greencheck IVI \redcross Ethics \redcross V2X \redcross Trust\newline
    \redcross Acceptance \greencheck Safety \redcross AI\newline
    \greencheck Disability needs \redcross Autonomy\newline
    \redcross Mode Adaptation \greencheck User Exp.\\

\hline
\addlinespace
\textbf{Ours** }&  & 
    \textbf{HMI analysis for diverse disability groups}\newline
    \textbf{Addresses Ethical and regulatory issues}\newline
    \textbf{Emerging AI features, multi-modal UX}\newline
    \textbf{Covers NDRA, mode adaptation, transitions, and awareness}\newline
    \textbf{Requirements for Trust, AV acceptance}\newline
    & 
    \textbf{N/A}
    & 
    \greencheck \textbf{IVI} \greencheck \textbf{Ethics} \greencheck \textbf{V2X} \greencheck \textbf{Trust}\newline
    \greencheck \textbf{Acceptance} \greencheck \textbf{Safety} \greencheck\textbf{AI}\newline
    \greencheck \textbf{Disability needs } \greencheck \textbf{Autonomy}\newline
    \greencheck \textbf{Mode Adaptation} \greencheck \textbf{User Exp.}\\

\addlinespace
\hline

\end{tabularx}
\end{table*}

\section{Recent Innovations and Accessibility}
This section highlights recent technological advancements with a direct impact or significant potential opportunities to enhance accessibility. As AVs transition from concept to reality, they hold the potential to revolutionize mobility for individuals with disabilities and older adults. Disabled /older users generally face difficulties in independent mobility, such as inadequate public transportation and reliance on caregivers. AVs help address these limitations by paving the way for safer, more convenient, and universally accessible transportation while potentially accommodating unique mobility needs \cite{krasniqi2016use}. However, eliminating this need for a driver also requires higher levels of automation. Thus, it's imperative to incorporate advanced technologies such as voice commands, accessibility features, and adaptive navigation systems to facilitate intuitive user interaction. 

Additionally, the other contributors are emerging assistive technologies outside the AV industry. Wearables like Ray-Ban's smart sunglasses offer significant potential for integration with AV systems. These glasses provide features such as haptic feedback for visually impaired individuals, bone-conduction speakers for those with hearing disabilities, \& AI-based voice commands to assist users with disabilities. By integrating these technologies into AV systems, situational awareness for blind individuals can be significantly enhanced. These devices provide additional information about the environment \& surrounding objects for the users \cite{zhang2024virtual}. Moreover, they are highly personalizable and can be configured per specific disability needs. For instance, blind individuals can focus more on audio and haptic interaction and some handy gestures like double tap to get quick feedback about the surroundings or any specific vehicle settings, and individuals with hearing disability can focus more on audio rendering through bone conduction with simple gesture activation. Similarly, cognitive disability users can generate gestures to activate easy help settings to remind them of specific features of the vehicle's voice activation to assist them with specific tasks like controlling Air conditioners, for instance. Wearable glasses are still in progress but have much to offer for accessibility.

Similar efforts in wearables can be seen with Apple Watches, which uses AI to combine motion sensors and optical heart rate monitor data to detect sign language for individuals with communicating disability. Their Assistive touch feature uses these sensor readings to support users with limited mobility with upper body limb differences. It allows them to control their watch without touching the screen or controls. Detecting subtle muscle movements and tendon activity allows users to control the watch with hand gestures, such as clenching a fist or pinching a thumb and index finger. This feature helps them perform tasks such as answering calls, controlling the onscreen pointer and accessing other device features\cite{apple2021assistivetouch}. Wearable devices can greatly help users engage and disengage in NDRA, serve for smooth handover and avoid mode confusion in critical scenarios.


\begin{figure*}[ht!]
    \centering
    \includegraphics[width=1\linewidth]{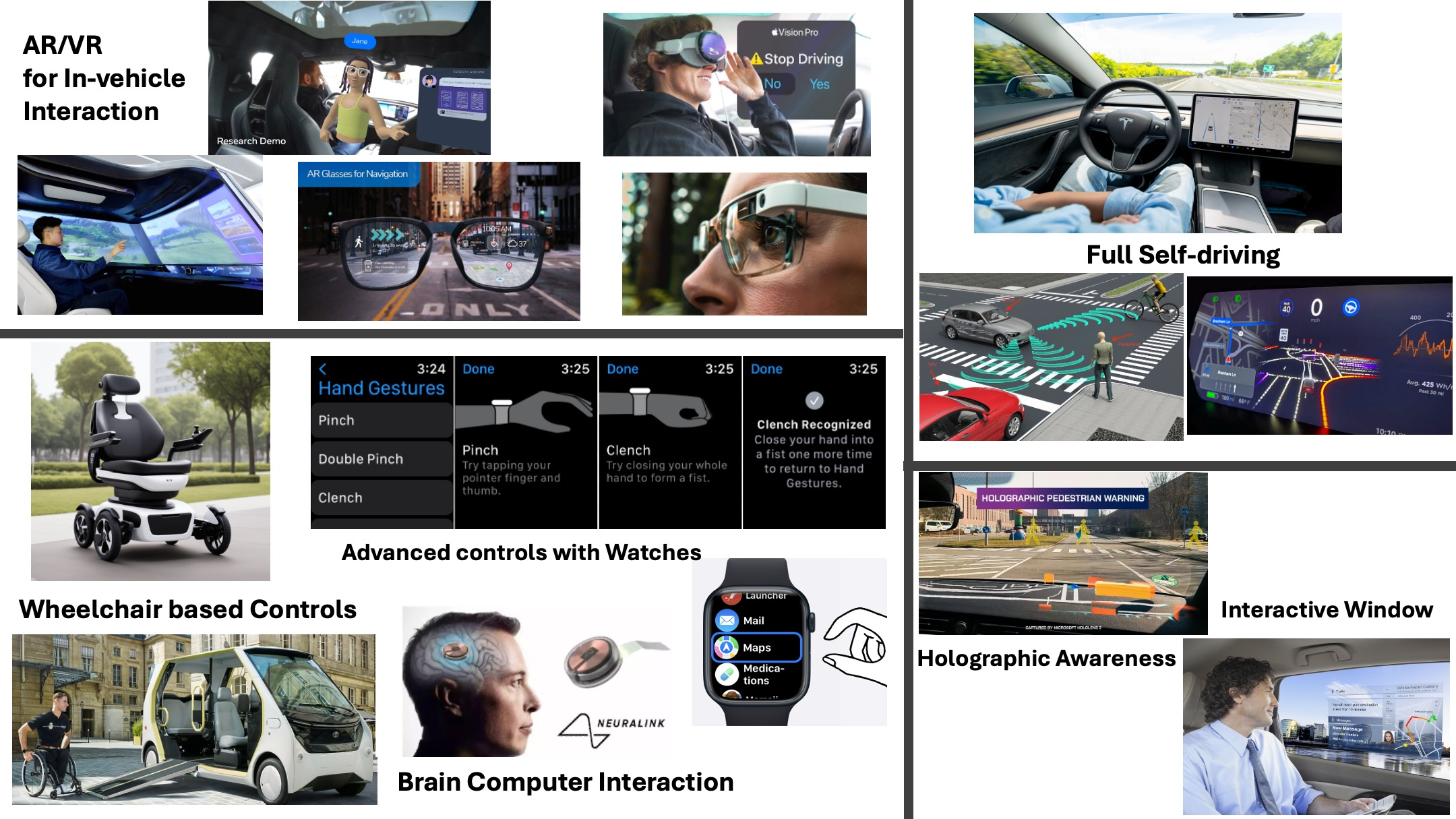}
    \caption{Current and Future Technologies that can be advanced to facilitate accessibility}
    \label{fig:current-accessible-technologies}
\end{figure*}

\begin{table*}[!htbp]
\fontsize{7.5}{9}\selectfont
\caption{Accessibility features of current wearable and potential integrations}
\label{table:wearables}
\begin{tabularx}{\textwidth}{p{0.06\textwidth}p{0.06\textwidth}p{0.06\textwidth}p{0.35\textwidth}p{0.37\textwidth}}

\hline
\textbf{Company} & \textbf{Devices} & \textbf{Category} & \textbf{Accessibility Highlights} & \textbf{Potential Integrations} \\
\hline

\addlinespace

Apple & 
    \textbf{Vision Pro}\newline
    \newline
    \newline
    \newline
    \newline
    \newline
    \newline
    \newline
    \newline
    \newline
    \newline
    \newline
    
    \textbf{Apple Watch}
    & 
    MR
    \newline
    \newline
    \newline
    \newline
    \newline
    \newline
    \newline
    \newline
    \newline
    \newline
    \newline
    \newline
    \newline
    Watch
    &

    \boxhl{green}{\textbf{Appearance Control} (Text Size, Bold Text, Brightness, Two-handed Window Zoom, Zoom Controller - Zoom Region, Use Crown to Zoom), \textbf{Interaction Controls }(Keyboard Shortcuts, Speak Selection, Eye Control(Both, Left Right), Pointer Control, Head Control, Wrist Control, Index Finger Control), \textbf{Situational Awareness}, \textbf{Voice \& Audio Feedback} (VoiceOver, Speak Screen, Speak Words While Typing, Audio Descriptions), \& \textbf{Motion \& Accessibility Adjustments} (Reduce Motion)} \newline
    
    \boxhl{red}{\textbf{Appearance Control }(Text Size, Bold Text, Brightness, Always-On Display), \textbf{Interaction Controls }(Digital Crown, Side Button, Touchscreen Gestures, Scribble, QuickPath Keyboard, Siri Commands), \textbf{Health \& Fitness }(Heart Rate Monitoring, Blood Oxygen Levels, ECG App, Sleep Tracking, Activity Rings, Workout Detection, Fall Detection), \textbf{Awareness Features }(Noise App, Handwashing Timer, Mindfulness App), \textbf{Voice \& Audio Feedback }(VoiceOver, Speak Screen, Haptic Alerts), \& \textbf{Motion \& Accessibility Adjustments}(Reduce Motion, Zoom, Taptic Time).}
    &
    \boxhl{green}{
    Navigation Info. (AR overlays for Route progress), Entertainment \& Productivity (Immersive media, Multi-screen for streaming \& work), Safety Awareness (Visual hazard alerts, AR Surroundings view, Audio descriptions for BVI), Personalized Controls \& Accessibility (Hand/eye tracking for climate \& lighting, Cloud-stored accessibility profiles for automatic adjustments), Social Interaction (Virtual meetings, Shared AR for collaboration), Assistance \& Guidance (vehicle features tutorials, AI-based cognitive assistance for reminders, emergency guidance), Sign Language \& Gesture Support (AR displays for real-time sign language translation, Simple gestures to signal emergencies), AR-Guided Assistance (AR overlays, voice guidance to help locate in-car objects for BVI)}. \newline
    
    \boxhl{red}{Watch enabled sign language translation, Haptic vehicle notification, AR-guided assistance (Vision Pro integration for locating features \& objects), Emergency Gestures (predefined gestures to send emergency alerts to vehicle), Health \& Wellness (Biometric monitoring, AR-guided relaxation), Personalized Accessibility Profiles, Real-Time Health Integration (adjusting to in-car conditions based on health metrics like stress or heart rate).}\\
    \addlinespace
    
Meta &
    \textbf{Orion} 
    \newline
    (Proto*)
    \newline
    \newline
    \newline
    \newline
    \newline
    \textbf{ Rayban }\newline
    \textbf{Smartglass}
    \newline
    \newline
    \newline
    \newline
    \newline
    \newline
    \newline
    \newline
    Quest
    &
    AR
    \newline
    \newline
    \newline
    \newline
    \newline
    \newline
    \newline
    Smartglass
    \newline
    \newline
    \newline
    \newline
    \newline
    \newline
    \newline
    \newline
    VR
    &
    \boxhl{blue}{\textbf{Display Technology }(Micro LED projectors, 70-degree field of view); \textbf{Interaction Controls }(voice commands, eye tracking, hand gestures, EMG wristband); \textbf{Design \& Build }(lightweight magnesium frames, under 100 grams); \& \textbf{Connectivity }(wireless compute puck for processing)}
    
    \boxhl{blue}{\textbf{Interaction Controls} (Touch controls on the frame, Voice commands with “Hey Meta”), \textbf{Awareness Features} (LED privacy indicator light when recording), \textbf{Voice \& Audio Feedback }(Built-in speakers for audio feedback, Five microphones for enhanced voice input, AI-based real-time language translation between English, Spanish, French, \& Italian)}\newline
    
    \boxhl{gray}{
    \textbf{Appearance Control }(Adjustable text size, Color correction for Deuteranomaly, Protanomaly, Tritanomaly), \textbf{Interaction Controls} (Voice commands via “Hey Meta,” Hand tracking, Customizable controller settings), \textbf{Awareness Features} (“Raise View” to simulate standing perspective for seated users), \textbf{Voice \& Audio Feedback (Audio balance adjustments for left \& right channels)}, \textbf{Motion \& Accessibility Adjustments}(“Adjust Height” feature for seated VR experience, Color correction for vision deficiencies).}\newline
    &

    \boxhl{blue}{Sign language translation, AR Assistance (locate seatbelt, controls etc.), Haptic Feedback (Alerts for navigation, hazards, \& emergencies), Bone conduction audio for deaf people, Emergency Gestures, Cognitive Assistance (AI-powered reminders or calming interactions for passengers with cognitive disabilities), Personalized Accessibility Profiles, Environment Awareness (AR overlays showing safe exits, nearby landmarks, or live traffic context for passengers with disabilities), Emergency Moral Support}
    \newline
    \newline
    \newline
    \boxhl{gray}{Immersive Accessibility(Real-time VR sign language translation, Virtual overlays for controls), Haptic alerts, Audio Enhancements(Spatial/Bone Conduction audio for immersive guidance), Gesture Recognition(for VR commands or signaling needs), Cognitive Support(AI-driven tools for reminders \& calming VR experiences), Personalized VR Settings(User-specific profiles for interface customization, VR based Situational Awareness, Emergency Assistance(Tutorials \& alerts integrated into VR for safety)}
    \\
Microsoft &
    \textbf{HoloLens}
    & 
    MR
    & 
    \boxhl{yellow}{\textbf{Interaction Controls}(Hand tracking, Voice commands for hologram manipulation), \textbf{Voice \& Audio Feedback }(Spatial audio cues for contextual information), \textbf{Ergonomics}(Ergonomic design with lightweight frame \& adjustable fit for comfort).}
    &
    \boxhl{yellow}{\textbf{Eye-Tracking for Navigation Control} (Gaze-based control of in-car interfaces for users with limited mobility), 3D Voice Commands(Spatially aware voice command recognition/user identification to facilitate use by mixed impairment groups), Hologram for Situational Awareness(outside pedestrian, vehicles etc.), Shared AR for multi-passenger accessibility, Live remote assistance etc.}
    \\

\addlinespace
Neuralink & 
    \textbf{N1 Implant}
    &
    BCI
    &
    \boxhl{purple}{\textbf{Interaction Controls }(Control of digital devices \& robotic limbs through thought, Enabling autonomy for individuals with paralysis), \textbf{Voice \& Audio Feedback }(Direct brain-to-device communication for seamless interaction), \textbf{Motion \& Accessibility Adjustments }(Restoration of motor functions by bypassing damaged neural pathways).}
    &
 
    \boxhl{purple}{Neural Control(Thought-based vehicle interaction), Accessibility control for physical or speech-impairment; eliminates reliance on voice \& gestures), Helth/Emotional Monitoring( Detect stress, seizures, fatigue, anxiety; adapts environment, issues alerts), Sensory Enhancements(Neural feedback for BVI users; real-time surrounding awareness), Adaptive Interfaces(AI-tailored controls based on cognitive patterns \& mental state), Collaborative/multi-user control}
    \\

\addlinespace
Google &
    \textbf{Google Glass}
    &
    Smartglass
    &
    \boxhl{orange}{\textbf{Interaction Controls }(Voice commands for hands-free operation, Touchpad gestures for navigation, Head gestures for display activation), \textbf{Voice \& Audio Feedback}(Bone conduction transducer for audio feedback without obstructing ambient sounds)}
    &
    \boxhl{orange}{Emergency vehicle control gesture recognition, head tilt motion for selecting options, advanced surrounding awareness using bone conduction audio, vehicle mode control etc.}\\
\addlinespace
\hline
\end{tabularx}
\end{table*}

Brain-computer interfaces (BCIs) such as Neuralink are yet another wearable transforming accessibility paradigm \cite{jawad2021engineering}. BCIs allow individuals to control systems, including vehicles, using only brain activity. While these technologies are in the early stages, they provide immense potential to eliminate the need for physical interaction with HMIs and smooth accessibility for individuals with severe motor impairments. Companies like Synchron and Paradromics also explore non-invasive(no surgery required) and semi-invasive BCI solutions that could further facilitate autonomous vehicle control. Such developments could eventually allow AVs to be operated solely through brain signals, removing barriers for nearly all disability types \cite{pisarchik2019novel}. However, fully integrating BCI technology with AV systems remains an open challenge, particularly ensuring reliability and safety during long-term deployments. These technologies reshape the future of wearable devices and especially serve great applications in in-vehicle interaction.

\subsection{Inclusive HMIs} 

Inclusive HMI(I-HMIs) refers to systems designed to ensure equitable access, usability and interaction for all individuals regardless of age, physical, or sensory and cognitive abilities. I-HMI essentially helps bridge the gap between technology and users with diverse needs, enabling seamless interaction through adaptable and intuitive design principles. Some of the key elements of a well-designed I-HMI include 
\begin{itemize}
    \item \textbf{Flexibility and Personalization}: Interfaces should be adaptive to individual needs, such as modifying control layouts or interaction styles based on their profiles. 
    \item \textbf{Error Tolerance}: Features like predictive text, undo options, and simplified controls that reduce cognitive load and minimize user errors.
    \item \textbf{Accessibility Integration}: Compliance with global accessibility standards(e.g., WCAG, ADA etc.) ensures that interfaces can cater to a broad range of impairments. 
\end{itemize}
 I-HMI can be used for any modern-day application; however, for our review, we focus on I-HMIs for in-vehicle interaction. However, we might also derive implementations from other applications that might be a potential implementation for in-vehicle use. 

Current in-vehicle HMIs integrate multi-modal interfaces such as voice commands, touchscreens, and haptic feedback, which improve usability for individuals with physical or sensory disabilities. For example, adaptive voice-activated systems enable hands-free operation, while tactile interfaces provide feedback for visually impaired users \cite{yan2022implications}. Meanwhile, recent research highlights the integration of augmented reality (AR) into HMIs, enabling visually impaired users to interpret their surroundings using real-time sensory data. Such advancements make vehicles more intuitive and accessible \cite{real2021ves}. Some of the most popular current and potential future Accessible technologies are demonstrated in \ref{fig:current-accessible-technologies}.

\subsection{Advancements in Wearables and their Potential in in-vehicle Interaction}
The evolution of wearable technology in the last decade has significantly expanded its applicability in various domains, including vehicles. Modern wearables such as smartwatches, augmented reality(AR) glasses, virtual reality(VR) headsets and brain-computer interfaces provide innovative ways for users to engage with vehicle systems, enhancing accessibility, convenience and safety. Innovations in AI-based hand gesture recognition, as demonstrated in the Apple Watch, are enhancing accessible interactions in complex sign-language translation. These systems have the potential to interpret hand movements to control navigation, entertainment systems, or vehicle settings, which is suitable for individuals with speech impairments \cite{mishra2021authorized}. Gesture control combined with haptic feedback can further improve and facilitate users with visual impairments by providing a physical sense of confirmation for action execution. These technologies can provide audio-visual alerts, situational updates, and emergency notifications, all enhancing interaction and accessibility for a broader user base.

Google Glass was the first consumer-oriented augmented reality (AR) device, launched in 2013 as a wearable glasses-based platform, paving the way for subsequent advancements in AR technology. Ever since, we've seen significant innovations in AR/VR. Some of the new mixed reality technologies, like Apple Vision Pro, have already incorporated significantly accessible interaction mechanisms such as eye-based controls with either or both of the eyes, speech selection, wrist control, head control and several personalizable accessibility adjustments. Microsoft Hololens is another such instance that incorporates holographic interactions. Google Glasses also initiated technologies such as bone-conduction audio, which is great for older users or users with significant hearing loss. These technologies, however, not currently incorporated in some of these most popular mixed reality headsets, might serve as an additional accessibility aid in the future. Above all, we've seen substantial innovation in Brain-Computer interfaces, which facilitate control by thought, meaning users can interact with digital interfaces by just thinking. Neuralink N1 is one such instance that has been successfully implanted and has shown great results in user interaction and can significantly boost the interaction capabilities of individuals with paralysis or any other significant physical limitations. These technologies, though in their current form, are invasive(requiring surgery); there is a great possibility that these will one day be used as a regular wearable without surgery. This will solve almost all interaction challenges, leading to a universally inclusive interaction mechanism that can be used even by users with no disability. We highlight recent innovations, their current functionalities, and their potential in-vehicle use in \ref{table:wearables}.

\subsection{Advancements in Vehicle Accessibility}
Major companies are driving innovations in vehicle interaction aiming to enhance safety and user experience. Companies like Tesla are advancing autonomous vehicle technology with their Full Self-Driving (FSD) software\cite{tesla_autopilot}, enabling hands-free travel for disabled users\cite{nordhoff2023mis}. While achieving full Level 5 autonomy may take time, we are closer than ever to experiencing these capabilities. Tesla's voice commands and app-based controls enable individuals with limited mobility or physical impairments to operate features such as opening doors, adjusting settings, and initiating driving modes without manual intervention. One of its popular Summon features allows users to retrieve the car autonomously in areas like parking, greatly reducing the dependency on others. The BMW iDrive system's new natural language processing accommodates users with varying technological familiarity, allowing intuitive voice commands\cite{bmw_idrive}. Their gesture recognition provides an alternate input method for users with physical impairments, reducing reliance on traditional touch buttons. The Mercedes-Benz MBUX's AI-powered voice assistant adapts to individual speech patterns, promoting inclusivity for users with speech impairments or accents. Augmented reality navigation simplifies complex driving scenarios, offering visual overlays that assist those with cognitive or visual impairments. Google's Waymo's ride-hailing services are designed to be accessible or passengers with disabilities. The app provides customization options for assistance during rides, such as wheelchair-loading support, and ensures seamless communication through intuitive user interfaces and visual cues\cite{waymo_support}. Rivian and Ford have incorporated similar features in the form of automated lift systems and modular seat adjustments that accommodate users with wheelchairs. Modular designs allow vehicle interiors to be customized based on user-specific needs, thus enabling greater flexibility for individuals with physical impairments\cite{zhang2020modular}. AI-assisted navigation systems can also predict mobility patterns and automatically adjust seat orientation or height to optimize accessibility \cite{singh2024ai}. Ford's features also facilitate V2X communication that enhances accessibility by enabling real-time traffic updates and alerts, improving safety for deaf or cognitively challenged drivers. The BlueCruise hand-free system with it's auto-drive empowers drivers with limited physical mobility to travel independently. 

Vehicle Companies also highly prioritize using Augmented Reality in the form of Heads-up displays and more immersive and large front displays. Some major ones include the BMW Panoramic Vision\cite{bmw_panoramic_vision}, which offers a full-width display projected onto the windscreen and critical driving information, navigation, and entertainment seamlessly integrated within the driver's line of sight. This also helps greatly enhance safety by reducing the need for drivers to glance away from the road. The Mercedes EQS True HUD facilitates AR overlays for navigation and projects turn-by-turn directions directly onto the windshield while also displaying key information such as speed, lane-keeping assistance and collision warnings in the driver's view\cite{mercedes_eqs_hud}. This expansive projection area provides a more futuristic and immersive experience while facilitating more room for personalization. Mercedes-Benz MBUX hyperscreen is another example with a massive curved display spanning the dashboard, integrating multiple screens into a single panel\cite{mbux_hyperscreen}. This AI-powered screen personalizes content and suggestions based on user habits and preferences while facilitating a passenger display for entertainment and controls, ensuring convenience for all occupants. They also facilitate haptic feedback and intuitive touch controls for easy user interaction. 

\begin{figure*}[ht!]
    \centering
    \includegraphics[width=1\linewidth]{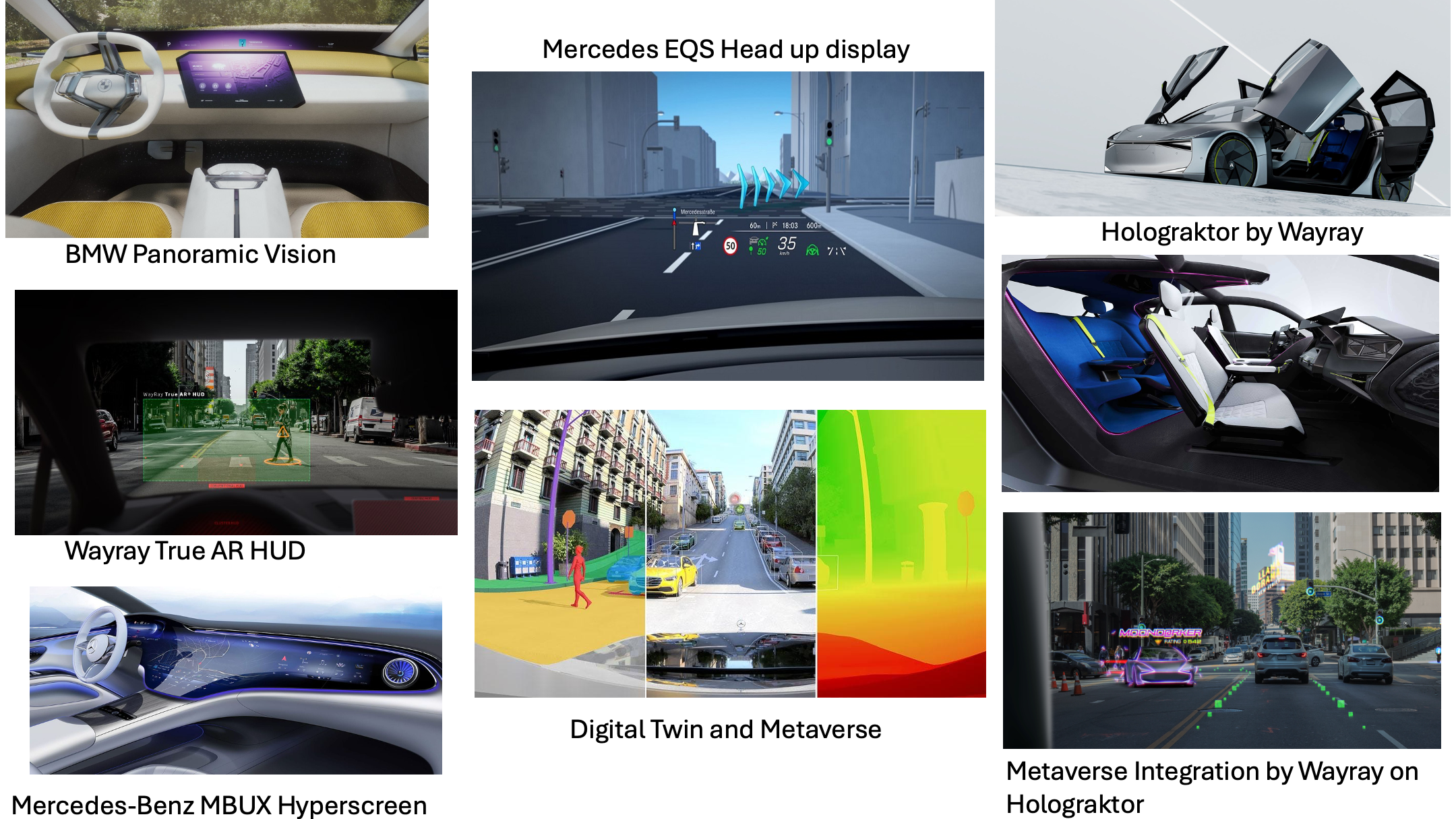}
    \caption{Innovative automotive display/visualization technologies, featuring BMW Panoramic Vision, Mercedes EQS Head-Up Display, Wayray True AR HUD, Mercedes-Benz MBUX Hyperscreen, Wayray's Holograktor, Digital Twin and Metaverse applications, and Metaverse integration by Wayray on Holograktor. Translation to VR opens up endless accessibility opportunities.}
    \label{fig:vehicle-ar}
\end{figure*}

WayRay, a Swiss deep-tech company founded in 2012, specializes in augmented reality (AR) and holographic display technologies for vehicles. It is one of the pioneers in True AR HUDs. It develops systems that project holographic 3D imagery onto windshields or other surfaces without additional wearables such as glasses\cite{wayray}. Some of its proposed holographic solutions seamlessly integrate elements of the metaverse into vehicle displays, delivering a cutting-edge and immersive interaction experience that defines the future of in-car technology. Its Holograktor, also shown in figure \ref{fig:vehicle-ar}, features holographic AR, which projects navigation, entertainment and contextual information in a more immersive manner in both its windshield and side windows for all occupants\cite{holograktor}. Their focus with this prototype is on ride-hailing and creating a unique experience for rear-seat passengers. These technologies highlight the leap towards more futuristic user interaction that can be a great accessibility benefit for disabled users as well. The above-explained display technologies are visualized in figure \ref{fig:vehicle-ar} and \ref{fig:current-accessible-technologies}.

One major challenge with these advanced technologies is their increased cost, making them less accessible and affordable for disabled user groups who could benefit the most. Thus, the cost is another factor that makes these technologies inaccessible despite the added accessibility features. 

\subsection{Data-Driven Personalization and User Adaptation}
AI-based personalization is an essential element in improving accessibility in AVs. By analyzing user behavior, past adjustments and preferences, AVs can dynamically adjust settings such as seat position, climate control, and HMI configurations. For instance, systems can prioritize navigation information or suppress non-essential alerts for cognitively impaired users \cite{schrum2024maveric}. Adapting machine learning algorithms \cite{boroujeni2024comprehensive} to predict user needs and deliver proactive assistance provides a significant advancement in AVs. These models allow vehicles to adapt dynamically in real-time situations to provide a seamless experience for users with diverse abilities. Real-time biometric sensors, paired wearables, and gaze-tracking devices can further enhance personalization. For instance, eye-tracking technology can monitor user engagement levels to mitigate issues like mode confusion or prolonged non-driving-related activities(NDRA). Additionally, AI-powered smart assistants can proactively set reminders for users, addressing memory impairments or alerting individuals of unsafe behaviors like disengagement during critical driving modes. By combining wearable data with AI models, AVs can adapt their functionalities to anticipate user needs and offer tailored solutions for individuals with cognitive or sensory impairments.

\subsection{Industry Collaboration and Policy Frameworks}
Advancements in accessibility are achieved by technological innovation and collaborative efforts across the industry \cite{taeihagh2019governing}. Partnerships between technology companies, disability advocacy groups, and policymakers help create standardized guidelines for accessible AV design. The National Science Foundation and other organizations have been instrumental in funding research initiatives focused on inclusivity. Regulatory frameworks that mandate universal design principles are crucial for widespread adoption. For instance, policies incentivizing manufacturers to include accessibility features can accelerate progress in this domain. Collaboration also enables identifying specific user needs through inclusive research practices, such as participatory design, ensuring that the voices of individuals with disabilities are integrated into the development process. Industry-wide adherence to universal design standards fosters innovation and levels the playing field for smaller companies aiming to create accessible technologies \cite{leon2022industry}.

\section{Vehicle Inclusion and User Experiences}
In this section, we discuss the factors influencing the inclusion of in-vehicle systems in AVs. These systems, designed to enhance accessibility and usability for diverse users, address the limitations of traditional interfaces by providing adaptive and intuitive interaction. By catering to the needs of older adults and disabled users, inclusive in-vehicle systems help improve the driving experience and promote independence by ensuring universal access to autonomous transportation, thus fostering social equity and mobility for all.

\subsection{Usage of In-vehicle systems (IVS)}
\label{sec:ivs}
Implementing in-vehicle systems is generally widespread, but their ease of use for individuals with disabilities is not guaranteed. These technologies may also negatively affect older adults, who rely on them rather than seeking physical assistance or human interaction. This challenges the automotive user interfaces \cite{riener2016automotive} and demands incorporation for further accessibility assistance. User Adoption and Acceptance are affected by these factors as well, which in turn influence the deployment of AVs in our society. It was revealed that, on average, participants have about 25 percent of the most available technologies in their vehicle, of which they only use 70\% of those available systems regularly.
In contrast, the rest, 30\% , were mostly unused or least used ones \cite{stiegemeier2022really}. The navigation system and light assistant were the most used, while the parking pilot, traffic jam assistant, and parking spot finder were the least used. Drivers who do not use certain systems attribute it to the fact that they do not need the system or trust the specific system or technology. To reduce clutter and cognitive demand, these technologies can thus be incorporated in a customizable configuration that allows users to personalize their frequently used features in the main UI while hiding unused ones. It is also logical to add less hardware that is software scalable to provide all the desired controls and features without a significant increase in cost while also providing a less intricate user interface.

\subsection{Quality of Service(QOS)}
\begin{figure}
    \centering
    \includegraphics[width=0.7\linewidth]{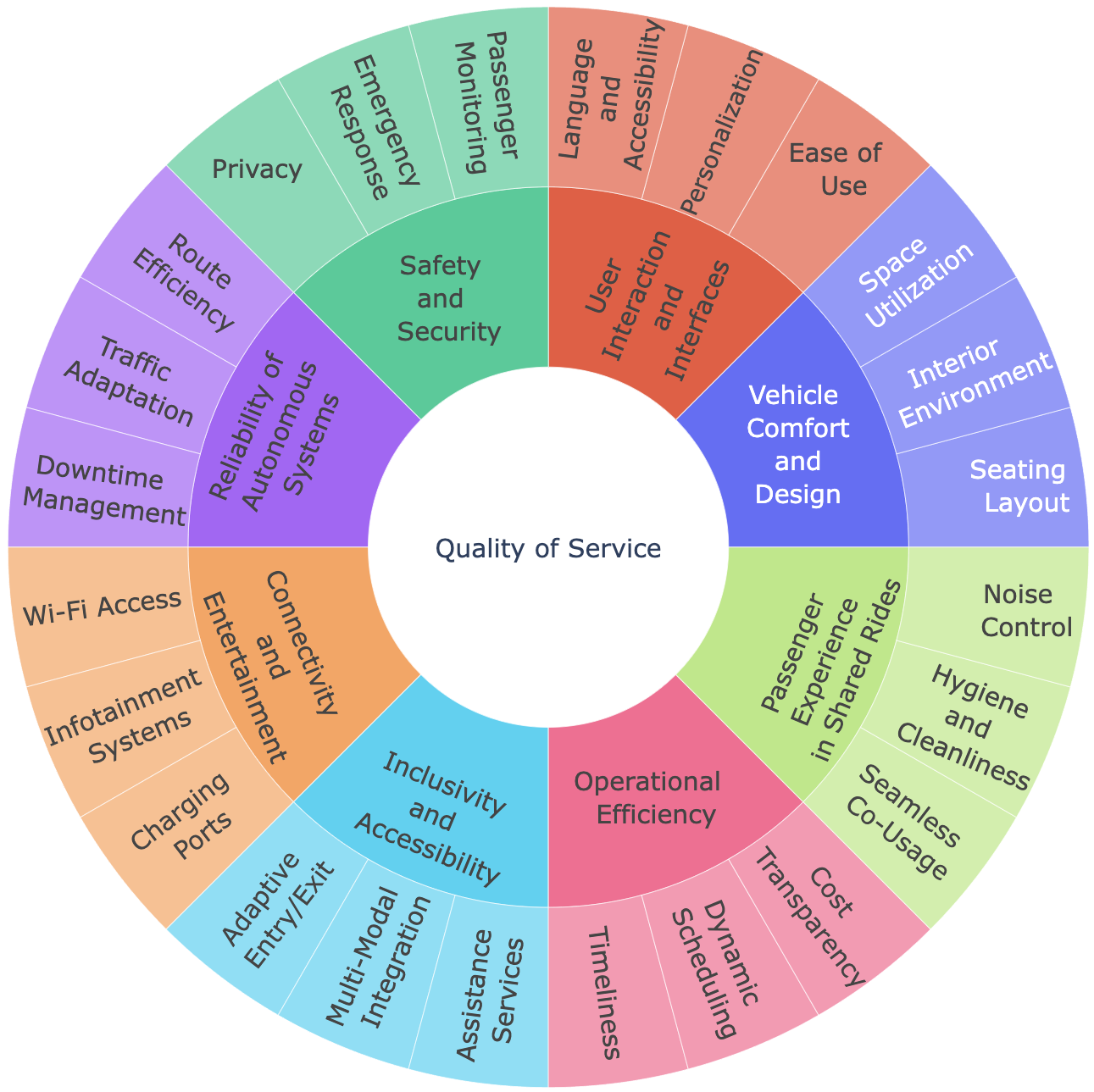}
    \caption{Factors affecting quality of service}
    \label{fig:qos}
\end{figure}

AVs are expected to become prevalent as mobility-as-a-service (MAAS) vehicles, primarily through ride-sharing services \cite{nazari2018shared}. QOS is the crucial aspect of user experience when using these services. Inspecting factors affecting overall service quality can be complex and even, in some cases, requires the developers and researchers to perform actual studies rather than ones that are simulator-based \cite{lee2022effect}, thus making the development cycle lengthy and expensive. However, based on the current research, the service quality during the traveling and drop-off phase is the significant factor that affects the overall satisfaction of the vehicle user \cite{lee2022effect}. Specifically, the reliability and robustness of the system, speed of travel, and kindness of the overall service were identified as crucial factors in the traveling stage. At the same time, accessibility, information, and communication were found to be essential in the drop-off stage. This can be even more pronounced for individuals with a disability, and the impact of these factors increases proportionally based on the inability's intensity. Physical well-being (including reducing motion sickness), peace of mind, aesthetics, social connectivity, proxemics, usability, association, and pleasure are essential while designing passenger experience in such vehicles \cite{diels2017designing}. Factors affecting the quality of services are shown in \ref{fig:qos}.

\subsection{User Preferences}

In AVs, user Preferences are shaped by a wide range of user needs, spanning comfort, safety, accessibility, and adaptability to individual lifestyles. These preferences are crucial to designing human-machine interfaces (HMI) that foster trust and satisfaction among diverse user demographics. Tailoring in-vehicle systems to accommodate specific user requirements facilitates the acceptance of autonomous technologies. It enhances the overall driving experience by addressing unique needs related to information accessibility, personalized driving styles, and system usability. By prioritizing these preferences, autonomous vehicle designs can provide a seamless and user-centered experience that meets the expectations of varied passenger groups, ensuring that all users—regardless of age, physical abilities, or familiarity with technology—feel supported and comfortable within AVs.

\subsubsection{Identification of User Needs}
To achieve broad acceptance of AVs across diverse user demographics, it is imperative to develop vehicular applications and user interfaces tailored to meet users' expectations. While such expectations are inherently subjective and vary with individuals and regional nuances in usage patterns \cite{wang2016drive}, it is essential to implement the necessary changes to meet all users' requirements. Determining these needs is a challenging process involving synthesizing input from various perspectives. It is critical to differentiate between genuine needs and mere wants and prioritize features that enhance usability. Expert interviews represent a particularly valuable source of information for this purpose. Various needs such as accessibility, personalization, user experience customization, space needed for various in-vehicle activities, information needs for presenting relevant information, and overall well-being was crucial for in-vehicle context \cite{lee2022eliciting}. Designing HMI also becomes complex due to the difficulty of requirements engineering, including poor representation or communication of user characteristics and needs. This is also true for individual interpretations by development team members, which could be error-prone and potentially biased by individual beliefs, leading to design deviation from what the user wants \cite{hallewell2022derivingPersonas}. It is imperative to consider both users and the experts in the design process to account for the errors caused by experts and the users themselves.

\subsubsection{Type and Detail of presented information}
The type and detail of information presented to the user play a vital role in optimally addressing the effectiveness of information transfer and their trust and user concerns. In the case of blind users, regarding the presentation and transmission of information, findings suggest that passengers of autonomous shuttles desired all three types of information (basic, technical, and supplementary) during their ride\cite{linnartz2022information}. The basic information would be similar to the information currently given by current public transportation systems, like the next destination and information such as stopping or departing. Technical information would explain how the shuttle works, and real-time sensor data of what the autonomous shuttle is detecting, to be shown by default. The supplementary information would be those that would be given in case of a problem with the shuttle and how it can improve the users' feeling of safety within the shuttle during the ride when such situations are encountered. Blind users also felt their concerns and apprehension being addressed when they were aware of their surroundings with the help of increased situational awareness and interaction, thus enhancing their intent to use these technologies \cite{brinkley2019open}.

\subsubsection{Personalization of driving styles}
Driving style plays a vital role in the perceived user experience of the system \cite{he2022battle}. This significantly impacts driving comfort and enhances driving enjoyment among younger and older participants \cite{hartwich2018driving}. Younger participants experienced decreased driving enjoyment, which could be due to the absence of secondary tasks, and also preferred a familiar AD style, while older drivers preferred an unfamiliar one. This might allow older adults to regain a driving style unaffected by age-related deficits. Also, older adults might want the vehicle speed to be slow, while younger, professional, and working-class users might prefer faster speeds \cite{clark2017age}. 

\section{HMI Design}
The human-machine interface (HMI) plays a critical role in the journey process, particularly for vulnerable user groups, despite being overshadowed by the popularity of AVs. Based on self-reported usability experiments with elderly adults, better-experienced usability of the HMI is positively correlated with cognitive abilities, particularly that of the working memory \cite{voinescu2020utility}. Additionally, trust in technology was positively linked to high usability scores for the HMI. Interaction, functionality, and contextual factors are the major dimensions that contribute to the design and evaluation of these interfaces when deployed commercially as a MaaS \cite{hallewell2022deriving}. The appealing experience of these interfaces is also crucial to persuade users to accept AV. There is no one-size-fits-all solution to designing an optimal HMI \cite{mathis2020creating} that accounts for the user experience. Thus, the fixed interfaces currently being deployed in commercial vehicles are inadequate. Instead, the system must adapt and be personalized to the user's needs. It is recommended to make these systems highly accessible at a bare minimum to ameliorate the issue of being inaccessible to some users instead. Including two or multiple of these HMIs with various multi-modal interaction mechanisms can also be a way to account for the needs that might not be fulfilled with the use of a single HMI \cite{hallewell2022derivingPersonas}, as in the absence of the human-driver to mediate support. One example in this context can be in making ingress and egress easier for disabled and older users by provisioning enough usability or providing multiple HMIs to provide minor corrections to the location to navigate to the exact location and prevent extra walking \cite{voinescu2020utility}. A specific HMI, in this case, can be helpful for them to fine-tune the location of the vehicle departure and arrival. Current Wearables and navigation systems\cite{bastola2023multi, wang2024motor} that assist with entry and exit navigation can be integrated with the vehicle's HMI to generate more precise localization and planning.

\subsubsection{eHMI and in-vehicle HMI integration}
Neglecting out-of-vehicle interaction while understanding the in-vehicle user experience might be problematic, as this increases the risk of accidents. With the increasing functionality of automation systems, communication needs and strategies with other Human road users change, and the passenger becomes less involved in the actual driving task. Considering many different HMI elements can be combined into an overall concept so that the requirements for the various automation levels and the role of the passenger can be met \cite{bengler2020hmi}. Examples include warning feedback in case of any unusual interaction with external users that may be dangerous similar to traffic accident detection systems\cite{bastola2023feedback}. Displaying what the seated vehicle eHMI currently displays might be a better way to inspect if the vehicle is providing the right information to the outside world and also be able to amend if not as desired. A recent study by Gelbal (2024) on vehicle-to-pedestrian (V2P) communication employed smartphone sensors and Bluetooth technology to relay pedestrian movements to drivers through warning systems which helped them with increased situational awareness. These integrations enhance pedestrian safety and situational awareness within the vehicle and improve the in-vehicle experience for disabled and older adults by creating alternative routes with fewer pedestrians. This reduces the anxiety, complexity and cognitive demands of manual control, making it easier by letting vehicles travel through less crowded environments. A similar approach could also include vehicle delays that allow pedestrians to cross the road, preventing vehicles from stopping too close to pedestrians.

\subsubsection{Heads-up Displays (HUD)}
HUDs have been one of the promising innovations in visualizing information in an autonomous vehicle. Along with these innovations comes the potential side effect of visual distraction and its significant impact on driver's cognition under different illumination levels, which might be critical for older adults, individuals with low cognitive abilities, and vision. Color luminance significantly affects visibility, with blue and purple colors having the lowest visibility \cite{zhong2022color}. White outlines significantly impact visibility, while grey outlines have no significant impact. Luminance and luminance contrast are critical design factors to improve cognitive efficiency and reduce visual fatigue. They thus are. Thus, they reference for designing AR-HUD interface characters in AVs and providing. Providing is also crucial for promoting trust and situational awareness when displaying certain information. It was also further shown that the HUD design affects the perception of external events and vehicle performance \cite{morra2019building}, thus demanding careful design consideration.

\begin{table*}
\renewcommand{\arraystretch}{1.5}
\small
\caption{Proposed Inclusive In-Vehicle Interaction Design Framework }
\centering
\begin{tabular}{|p{0.15\linewidth}|p{0.68\linewidth}|p{0.10\linewidth}|}
\hline
Interaction Phase & Design Considerations & Supporting Articles\\ \hline
Ingress 
& - Provision to arrival at the precise location using smartphone-based control for fine-tuning & \cite{voinescu2020utility, fink2021fully} \\
& - Handles pop up and seat adjust automatically to maximize space for easy ingress & \cite{wang2016drive, lee2022eliciting}\\
\hline

Interaction with the HMI
& - Smartphone integration: Customize UI based on user profile and media control. Show only what's needed  & \cite{fink2021fully, stiegemeier2022really}\\
& - Multi-modal interaction: haptics, audio, and touch to tailor interactions for user activities & \cite{farooq2014developing, zheng2022developing}\\
& - Provision easy and minimal UI to reduce cognitive and memory demand & \cite{voinescu2020utility, eimontaite2020impact}\\
\hline

Type and detail of presented Info
& - Situational awareness of surroundings, adaptable to turn on or off and adjusting verbosity and frequency of information delivered. Prioritize on 'why' vehicle takes certain actions & \cite{morra2019building, brinkley2019open}\\
& - Critical information about road users or emergencies that need the user to take over & \cite{bengler2020hmi}\\
& - HUD: Poor luminance, contrast, and colors can be distracting and mentally demanding & \cite{zhong2022color}\\
& - Prioritize map Graphics optimization more than font size to ease blind and elderly users & \cite{angeleska2022inclusive}\\
& - Information from other vehicle eHMIs or infrastructure embedded eHMI reduces arousal and lets the user be prepared for handover ahead of time, providing ample time for takeover & \cite{lingam2022ehmi}\\
& - Anthropomorphic agents improve conversational interaction, reducing loneliness and increasing perceived trust, pleasure, and companionship during the journey& \cite{large2019please}\\

\hline

Adaptation/
& - Let users elicit their needs and concerns and personalize driving experience accordingly & \cite{brinkley2020exploring, fernandes2020challenges}\\
Personalization & - Prioritize Personalized speed, level of kindness, and verbosity of the system interaction & \cite{lee2022effect}\\
& - Provision for Region-based adaptation for the HMI and user expectations & \cite{wang2016drive}\\
& - Adjust displayed information level (basic, supplementary, technical) based on user profile & \cite{linnartz2022information}\\
& - Profile-based driving styles among age groups, disabled users, and working professionals & \cite{hartwich2018driving, clark2017age}\\
\hline

Additional HMI 
& - Multiple HMIs for different functionalities. HMI that allows fine-tune egress location & \cite{hallewell2022derivingPersonas, mathis2020creating, voinescu2020utility}\\
Consideration & - Embed interfaces within seats to provide easy access & \cite{custodio2022change}\\
\hline

In-vehicle needs 
& - Physical well-being, Aesthetics, Social Connectivity, space needs for in-vehicle activities, the usability of system and pleasure of driving & \cite{lee2022eliciting, zheng2022developing, diels2017designing}\\
& - Let the users elicit their needs to personalize their driving experience. Older drivers and BVI are extremely sensitive to this and express concerns about whether their needs are addressed in the design & \cite{brinkley2020exploring, fernandes2020challenges}\\
\hline

Vehicle Takeover 
& - Takeover warning using multi-modal feedback & \cite{farooq2014developing, bengler2020hmi}\\
& - Notify takeover status using voice and text before and after with reasons to reduce workload and enhance positive attitude. Ease posture to regain control easily & \cite{li2019evaluation, brandenburg2019drivers, clark2019directability, you2018take}\\
& - Adjust the pre-takeover alarm period based on the age group. V2X communication to identify critical areas by communicating with infrastructure, enabling advance rider notifications.& \cite{li2019evaluation, kinnear2015battle, large2017design}\\
& - Additional assistance for quick driver re-engagement e.g. detecting pedal misapplication& \cite{inoue2014research, lu2022can} \\
\hline

Mode Confusion 
& - Provide enough mode information to avoid state anxiety and allow quick engagement. & \cite{lu2022can}\\
& - Do not rely on the steering wheel to detect mode confusion, embed gaze behavior detection & \cite{wilson2020driver, haghzare2022classifying}\\
& - Articulate driving mode, consider NDRA impact to adjust takeover warning delay&\cite{wolter2020human}\\
& - Displaying NDRA task availability time boosts system usability, reduces unfulfilled expectation frustration, and enables smooth system re-integration.& \cite{danner2020does}\\
\hline
\addlinespace
\end{tabular}
\label{tab:ProposedFramework}
\end{table*}

\subsection{Texts and Visuals}
Customizing regular text is a highly desirable feature in interface design. However, the importance of text within certain applications, often overlooked, can be equally significant. For instance, When displaying information to low-vision individuals using a regular in-vehicle display, it was found that regular font type and size are less of an issue than the generally available navigation map, which is mostly unclear, suggesting a need for optimizing map graphics and adjusting the map size\cite{angeleska2022inclusive}. 

\subsection{Multi-modal User Interaction}
Touch, haptic, and voice-based interfaces have great potential to make AVs more accessible to people with disabilities\cite{bastola2023multi, bastola2023feedback}. By providing an intuitive and easy-to-use interface, touch sensors can enable users with various disabilities to control and navigate functions from their driver's seat. The effectiveness and reliability of touch-based sensors also make them an appropriate choice for the automotive industry \cite{custodio2022change}. Similarly, voice-based interfaces allow for hands-free and distraction-free interaction with the vehicle's systems, improving safety and convenience for passengers. Using voice-based status feedback has resulted in greater situational awareness and overall positive impact. Results indicated that participants in the condition without audio notifications required the highest level of concentration, suggesting the increase in ease of use and overall journey experience \cite{eimontaite2020impact}. Integrating smartphones with the in-vehicle system can be one way to scale the personalization of user experience \cite{fink2021fully} and facilitate an additional HMI. Studies have claimed integrating smartphone-based apps (e.g., Waymo's integration) can help with easy information processing\cite{bastola2023llm} and solve many accessibility challenges that BVI individuals face. This approach will enable audio and haptic interaction capabilities for completing various tasks related to autonomous mobility and fit well into the critical need for broadening the applications of information access technologies. Haptic feedback improves users' primary task performance and adds an immersive experience to the in-vehicle interface\cite{farooq2014developing}. There are tremendous opportunities afforded by touchscreen-based smart devices that employ native multi-modal feedback mechanisms, for these can serve as a primary channel of haptic interaction and convey spatial information, such as graphical and non-textual information, which are inaccessible to current screen readers. In case of vehicle takeover, delivering driving state information and warnings, visual, haptic-tactile, and auditory signals are the suitable cues \cite{bengler2020hmi} that can be delivered using smartphones.

\subsection{Enhancing Accessibility through V2X}
Vehicle to Everything(V2X) is a communication technology that enables real-time data exchange between vehicles and various entities in the environment, including other vehicles(V2V), infrastructure(V2I), pedestrians(V2P) and networks(V2N). By leveraging wireless communication, V2X can help enhance road safety, traffic efficiency and overall driving experience by notifying users of information such as road closures, evacuation routes and hazardous conditions relayed through decentralized systems such as CCTV cameras\cite{bastola2024fedmil}.  Real-time exchanged information can be relayed through vehicle HMIs, improving situational awareness. For instance, a prototype was developed to display V2X safety alerts to drivers and passengers, capturing messages from radio communication between the vehicle's onboard unit and external sources. The system presents alerts such as emergency brake lights, forward collision warnings and red-light violation warnings, which can be particularly beneficial for drivers with disabilities when provided in a timely manner\cite{wallace2023hmi}. There is a possibility that V2X-enabled vehicles can receive priority at traffic signals, reducing travel time and complexity for drivers with mobility challenges \cite{ITS_America2024, USDOT2023}.

Additionally, V2X technology facilitated communication enables features like signal priority and preemption(prioritize emergency vehicles). These capabilities can be integrated into in-vehicle systems to provide real-time updates on traffic signals and road conditions, allowing drivers with disabilities to navigate more safely and efficiently\cite{USDOT2024}. 
Moreover, V2X technology supports applications that assist pedestrians with vision disabilities. Field tests of mobile applications utilizing V2X communication have shown that 83\% of participants felt safer when using the app compared to not using it\cite{Walker2022}.
One of the most critical benefits of Vehicle-to-Everything (V2X) technology is its ability to drastically reduce emergency response times which is essential in case the user in-vehicle is having difficulty gaining control. By enabling real-time communication between vehicles, infrastructure, and emergency services, critical information can be shared instantly in life-threatening situations. For instance, when a vehicle detects a crash or sudden loss of control, V2X can automatically alert nearby emergency services and transmit precise location data, traffic conditions, and even vehicle occupant details. In other terms, decentralized V2X roadside units (RSUs), such as CCTV cameras, can assist in detecting anomalous vehicle behavior or accidents and relaying this information to vehicles\cite{bastola2024fedmil}. This allows first responders to plan their routes more efficiently and arrive at the scene faster, potentially saving lives in scenarios where every second counts\cite{ITS_America2024, USDOT2023}.

\subsection{Role of AI in Advanced Interaction and Research}
Current AI-based interactions, though seemingly similar, are completely different than those in the past few years, like simple voice-based interaction, VR-based interaction, simple assistive agents etc. Advanced data-driven AI has recently been introduced with the advent of Large Language(LLMs) and Diffusion Models and has shown tremendous capabilities in realism, intelligence and reasoning. BCI is another major application that AI will advance. Generative AI models, including transformers and diffusion models, can significantly enhance BCI systems in the development phase by improving brain function understanding to guide researchers, rehabilitation, and actual usage once deployed. Various LLM deployments, such as ChatGPT and Grok, are highly capable interactive agents paired with recent hyper-realistic video generation models such as SORA; there is a great possibility to create hyper-realistic anthropomorphic agents with human-like voices, facial expressions, or gestures and can serve as more intuitive and relatable in-vehicle agents. By simulating human communication patterns and emotional cues with human-like personnel, these agents can help enhance user trust and comfort by helping them feel at ease. Another benefit of these advanced AI incorporations is improvement in communication clarity. These agents can convey complex information, such as navigation updates or safety alerts, in an engaging and easily understandable manner. Personalization is another benefit, as they can be customized based on tone, language, complexity, and feedback based on user preferences, driving conditions and accessibility needs. These agents can also facilitate smooth transitions between handovers as well. With carefully crafted prompts, they can help better guide users to avoid mode confusion and relieve the cognitive load, making AI much more approachable, trustworthy and supportive and aiding user acceptance and comfort. 
\cite{eldawlatly2024role} explores applications like data augmentation, signal enhancement and more complex pattern recognition that AI can enhance. BCI users show large variability in the usage patterns and have difficulty controlling the signal response known as BCI illiteracy. With Adaptive AI integrations such as in Neuralink \cite{zhang2020combination, musk2019integrated}, they can learn from previous data and help with BCI illiteracy by helping them gain control of the system. Moreover, the more these systems are deployed, much more capable system can be developed with distributed training by learning to account for variability across users, making the interaction easier for all user learning abilities. Ethical and privacy-related concerns are inherent to training with these data types and should be addressed in future studies once the technological deployment is more mature.

Speech to text applications are another major area where AI contributes. Given the current LLM enhancements, AI can generate video captions and enable voice control functionalities with much more precision. \cite{jain2022protosound}. Google has brought up significant steps in enhancing the accessibility here. The project Euphonia enables a personal speech model for users with atypical speech by collecting utterances from hundreds of people around the world\cite{GoogleResearch2022Euphonia, EqualEntry2023AIAccessibility}. The personalized model reduced the error rate significantly from 31\% to 4.6\% and, in many cases, outperformed human transcribers unfamiliar with a particular person's speech or communication style. A graphene-based wearable artificial throat has been developed to accurately recognize vocal patterns and generate realistic speech, especially for those with voice disorders or communication difficulties \cite{yang2023mixed}. 
Google also released an app, Project Relate, which provides a personalized interaction medium for these disability groups\cite{projectrelate}. Environmental sounds are another way to provide situational awareness, and ProtoSound \cite{jain2022protosound} is yet another effort that identifies sounds in the environment and displays the entity related to that sound. For instance if a dog is barking, a picture of a dog would be shown. They also facilitate the personalization aspect so that users can make a few recordings of sounds of their desired environment (like home or office, for instance; some generic sounds like a fire engine, for instance, are provided) and train the corresponding representations. Text filtering and text-to-speech are other applications that greatly benefit from AI. People who struggle due to dyslexia, lack of fluency or even low vision to interpret the text can have a personalized view of the text that they want to read with technologies like Android Reading App \cite{androidreadingmode}. It can also read out text with the most recent text-to-speech models with more expressive natural voices that are easy to understand. It can also help filter out unwanted content across the screen. Similar technology can have great potential to be integrated into vehicle features, allowing selections to be biased toward the user's history unless the user specifically targets areas on the screen.

AI has advanced substantially in understanding complex image representations. Visual assistant tools like Google Lookout\cite{googlelookout} and Microsoft's SeeingAI\cite{microsoftseeingai} convert camera-detected objects into audio information. They also serve as screen readers and can identify currency, read barcodes, and more. These tools are becoming more personalizable, allowing users to train them to recognize specific items in their homes by letting them train with their own captured images and labels.

\subsection{HMI Design Methodologies/Principles}
The design of the HMI is a crucial determinant of the prevailing user interaction trend that has continued for years. The proposed frameworks must also be flexible and generalizable to future configurations to establish an inclusive design culture. To account for accessibility, impairment-specific sensory substitutions should be made available to compensate for one's dysfunctional sensory input\cite{brinkley2024atlas}. Incorporating inclusive design principles in the HMI is thus essential to ensure that it accommodates the diverse needs and preferences of a wide range of users, regardless of their abilities. Lack of standardization in the design of automated functions may lead to confusion and human error \cite{carsten2019can}. One of the most considered HMI design methodologies includes participatory design. Researchers claim that, given the low-cost, low-impact characteristics of the project and few resources available, the participation of one co-designer is sufficient in the design process. It does not risk the over-design of the prototype \cite{huff2020participatory}. Some of the key considerations while using a participatory design are to make sure the environment is accessible, user design methods are adaptable as per the needs of the co-designer (select methods that require less or no visual proficiency), design solutions are in accessible formats, and a right number of co-designers are considered as increasing the number uselessly would not contribute much to the design. A viable approach for aiding HMI researchers to design a better HMI involves finding research gaps that contrast various elements of vehicle user, vehicle, target activities, and system Input/Outputs \cite{zheng2022developing}. The second would be to specify subjects, i.e., the passenger's personal information and scenario of vehicle usage. The third step involves specifying the target activities, which involves researchers clarifying the aspects of wellness they hope to solve through HMI design while considering passengers' needs and demands sufficiently. For the fourth step, specifying system interactivity involves systematic consideration of the input and output of the system and then comparing which combination of system IO can achieve the target activities more effectively. The final step is the design of HMI for autonomous wellness within the bounds of the elements mentioned above. These design principles can be leveraged to develop interfaces for older and disabled users. 

\subsubsection{HMI and Autonomous Driving Experience Evaluation}
Interaction evaluation is equally important to validate the designed HMI configuration. Self-report measures have been more effective for evaluating HMI designs for Level 3 Automated Driving and above \cite{forster2020self}. Self-report methods include asking users to report their thoughts, feelings, and experiences while interacting with an interface using rating scales, questionnaires, interviews, etc. A virtual reality simulator for HMI evaluation has also been found to enhance the ecological validity of the system \cite{voinescu2020utility} interaction. Studies suggest validating autonomous driving systems by gathering continuous, quantitative information from physiological signals during a virtual reality driving simulation can be a great way to understand the user's condition and gain valuable insights on the user experience \cite{morra2019building}. This methodology was also shown to aid in designing sensory subsystems by considering practical HMI constraints, further improving users' acceptance of autonomous driving systems. Another effective way to evaluate the HMI experience would be an ethnographic study, which provides in-depth insights into how users use the system and account for their concerns in a real-world scenario. Adding before-and-after ethnography to the Wizard of Oz experiment has been shown to yield unexpected insights grounded in human experience and expectations of everyday driving and commute \cite{osz2018building}. In this method, before and after interviews are conducted with study participants immediately before and following WOz testing sessions. This approach enables the participants to delve into their emotions (such as trust, mistrust, or anxiety) and their distinctive driving habits and commuting patterns and examine how they relate to their overall experience during the test.
    
\section{Driver Adaptation and Handover Mechanisms}
Creating a suitable user adaptation system has been one of the major challenges in accounting for anxiety during travel, situational awareness, and the driver's adaptive role. The interaction must meet user needs while being adaptive enough to convey all information and allow quick driver re-engagement. State anxiety has a significant negative effect on trust, situational awareness, and role adaptation \cite{lu2022can} if not handled carefully. The transitional phase of vehicle automation presents critical challenges, where human drivers have a truncated yet crucial role in monitoring and supervising vehicle operations. However, numerous challenges remain to overcome concerning the continued role of human drivers, including safety, trust, driver independence, failure management, third-party testing, and regulation of current and future vehicle automation technologies \cite{hancock2020challenges}. Research suggests considering a multidisciplinary approach to address the challenges of these technologies when considering the evolving roles of vehicles and users. Additionally, balancing competing priorities in the design of transportation is crucial. We present some crucial aspects of driver adaptation below:

\subsection{Engagement in Non-driving Related Activities (NDRA)}
One of the primary advantages of using AVs is the ability to engage in non-driving-related activities (NDRA). Reducing human control in AVs raises concerns, particularly regarding in-vehicle interaction. 
Therefore, the HMI must support users in NDRA and manual control modes by providing appropriate information and feedback. For example, displaying the duration of NDRA availability can enhance safety and facilitate timely re-engagement. Providing appropriate features during NDRA can help users have a much more entertaining ride. This can greatly help improve system usability and acceptance, as it reduces cognitive load(especially when engaging in NDRA) and makes interactions more purposeful\cite{danner2020does}. Additionally, effectively presenting this information using appropriate user feedback(for instance, added haptic feedback for hearing disability and next steps for cognitive) and images can alleviate frustration and allow for a smoother transition to NDRA. Drivers can engage with peace of mind, knowing they will be notified well before a required handover. However, the type of information and feedback might need to vary among different user groups; for instance, individuals with cognitive impairments may require not only a timer but also guidance on the next steps, while those with hearing impairments might benefit from haptic feedback to draw their attention to the display when audio cues are insufficient. This necessitates personalization according to user needs to ensure quick and effective re-engagement.

\subsection{Detecting mode confusion}
The use of partial automation on public highways has demonstrated increased acceptance and trust of autonomous vehicle technologies among riders and improved perceived safety. However, research indicates a potential safety issue related to decreased engagement and monitoring of the roadway compared to manual driving. Additionally, steering wheel sensors are unreliable in assessing driver engagement with the monitoring system, leading to more confusion and increased risk as drivers may engage in non-driving activities while still believing the vehicle is in control \cite{wilson2020driver}. With mode confusion, the driver is confused about the vehicle's current operating mode and, therefore, about their role. This is a serious concern for older individuals and individuals with low cognitive abilities who are more likely to have poor situational awareness of the system. One way to detect older drivers' mode confusion has been by inferring the driver's perceived AV mode using gaze behavior. Gaze behavior can be identified using a classification model pre-trained on eye-tracking data collected from the participants. The features could distinguish between the driving scenarios of automated and non-automated as perceived by the drivers \cite{haghzare2022classifying}. 

The level of information on driver takeover guidance also plays a vital role in takeover performance and mode confusion. HMI, which informs the driver of the status and reasons for the takeover (can be verbal), is found to facilitate good takeover performance, lower perceived workload and increased positive attitudes, thus being an optimal HMI interaction approach \cite{li2019evaluation}. Age differences can also significantly affect the driver's takeover performance. Compared to younger drivers, older drivers were found to take longer to switch back to the manual driving position after receiving the takeover request. They were also slower to make lane-changing decisions to overtake a stationary car ahead \cite{li2019evaluation}. Furthermore, even with this approach of providing takeover information ahead of time, older drivers had the highest resulting acceleration, steering wheel angle, and riskiest takeovers. Thus incorporating older drivers in the design process while considering their capabilities and needs contributes significantly to developing accessible takeover. It might be a good approach for these vehicles to communicate with intelligent infrastructure to update drivers with information on system limitations and concerning road and traffic conditions. This emphasizes building collaboration between the automated vehicle research community and the C-ITS (Cooperative Intelligent Transport System) \cite{li2019evaluation}.

\begin{table*}[ht]
\footnotesize
\renewcommand{\arraystretch}{1.5}
\caption{Additional Consideration for Overall AV Acceptance}
\centering
\begin{tabular}{|p{0.15\linewidth}|p{0.655\linewidth}|p{0.14\linewidth}|}
\hline
\textbf{Context}&\textbf{Proposed Solution}&\textbf{Supporting Articles}\\ 
\hline

\textbf{User Needs}
& - find research gaps in vehicle-user abilities, vehicle, target activities, and system I/O& \cite{zheng2022developing}\\
\textbf{Identification} & - Balance expert and user input in the design. Iterative user and expert design-validation works well& \cite{wang2016drive, hallewell2022derivingPersonas}\\
& - Participatory Design, Self-report measures & \cite{huff2020participatory, forster2020self, dong2024review}\\
& - VR simulator study, Gathering physiological data during VR simulation & \cite{voinescu2020utility, morra2019building}\\
& - Ethnographic study-  before-and-after WoZ study & \cite{osz2018building}\\
\hline

\textbf{AV Training}
& - Knowledge of AV operation can be a factor that hinders AV acceptance & \cite{stiegemeier2022really, park2024automated, li2024effect, sharma2022quantifying}\\
& - AR/VR training for takeover in critical situations leads to faster reaction time in real driving & \cite{sportillo2019road}\\
& - Providing an opportunity to take part in real autonomous rides can be even more effective in enhancing their control and operation of vehicle & \cite{classen2020older, large2019please}\\
\hline

\textbf{Trust and Affordability}
& - Trust directly affects perceived safety and driving comfort. Increase trust by marketing AV taxi services and providing early service experience to get the users used to & \cite{lee2022effect, hartwich2020passenger, fernandes2020challenges}\\
& - Take Affordability into account when building technologies for BVI and elderly who have relatively low income & \cite{bennett2020willingness, blindIncome}\\
\hline

\textbf{Design Standardization}
& - Government-level standardization promotes consistent design configurations across all services, facilitating ease of use and allowing users to access multiple services interchangeably& \cite{carsten2019can} \\

\hline
\end{tabular}
\label{tab:AdditionalConsideration}\\
\end{table*}

\subsection{Driver Handover Situation}
The transition from automated to manual driving can be a major issue related to human factors in AV usage. There is a need to communicate the active driving mode unambiguously, taking into account the impact of NDRA carried out by the user while driving to adjust the delay period of the warning for takeover \cite{wolter2020human}. Thus, the importance of monitoring the driver's attention and other road users not equipped with automated driving functions exists. The system additionally needs to emphasize the differences in infrastructure and inform the vehicle driver if manual driving is recommended in a specific location. Lower-level AVs might need more manual engagement, and thus, users should be prepared for re-engagement. Cognitive decline is one of the major factors leading to the inability to drive manual vehicles in older adults. The prevalence of dementia doubles every 20 years because of the aging \cite{dementiaStats}, and thus, even middle-aged users can face the consequences. Older drivers experience age-related declines in visual perception and cognitive functions but can compensate by driving more cautiously. However, they may have difficulty with distractions like mobile devices and navigation systems and may take longer to respond to hazards \cite{kinnear2015battle}. It was also revealed that participants might face age-related presbyopia (farsightedness), which requires them to wear reading glasses while engaging in in-vehicle activities. This can be even more pronounced when a user is required to take manual control of the vehicle quickly. Studies identified that even five seconds of emergency hand-over period was insufficient for older participants to remove and store their glasses safely \cite{large2017design}. Thus, it is also important to incorporate an adaptable system that warns the user ahead of time (adjusted per their conditions) to provide enough time to re-engage. AV riders were also found to prefer a two-step procedure using text and speech for communication rather than a one-step procedure to inform for takeover. It has also been demonstrated that the best takeover request interface received significantly higher user experience ratings than the worst \cite{brandenburg2019drivers}. It is also recommended to provide additional assistance in reversing, parking, and pedal misapplication and the ability of the system to override manual control like autonomous braking in cases when the driver takes an accidental step, which also needs to be handled by the driver adaptation system \cite{inoue2014research}.

\subsubsection{Use of AR and VR}
Training users can be an effective way for them to successfully perform take-over in real driving scenarios without any emergency stops. Training using AR/ VR programs resulted in faster reaction times than the video tutorial. It provided a better sense of immersion and isolation, which helped the participants better familiarize themselves with the vehicle and driving situations \cite{sportillo2019road}. Thus such technologies have been proven beneficial for older adults in better understanding the usage of AV and, most importantly, enhancing trust in autonomous systems. 

\subsubsection{Communication with other Road Users}
Although in-vehicle user interaction is typically the focus of attention, handover situations are critically influenced by events that occur in the outside world. In these instances, information is conveyed from the environment to the driver or vehicle operator, particularly when other vehicles are present on the road. Utilizing eHMI in other vehicles could facilitate the efficient communication of information regarding status or intent to vulnerable road users. These interfaces may be located within a vehicle or in the surrounding infrastructure, enabling them to communicate with other vehicles continuously. Infrastructure-embedded eHMI positively affected lower arousal and earlier slowing down of the vehicle and provided the driver ample time to take manual control if necessary \cite{lingam2022ehmi}. These eHMI systems contributed equally to the participants' crossing decision compared to that in the vehicle. Two significant components that contribute to the effectiveness of this approach are the separation of information and its intended recipient and the non-distracting, easy cognitive processing of the information. If not properly addressed, these two factors may create cognitive overload and confusion among older adults or individuals with lower cognitive abilities taking control of the vehicle. Thus, further research is still required.

\subsubsection{Voice-based Guidance}
As discussed earlier, the driver's eye gaze is vital in the driver's handover and manual interaction. Vocal guidance is proven to guide visual attention effectively. However, user-based customization is required due to variability in visual information stream utilization \cite{clark2019directability}. Current AV designs may not address unique handover interaction requirements, as evidenced by minimal center console gazing during manual driving. Furthermore, when evaluated in level three and four vehicles in a free-form and pre-defined condition in a dual-controlled driving simulation with two human drivers, it was identified that the drivers were open to information transfer and preferred interactive questioning and checklists \cite{clark2019conditionally}. Similarly, when driver performance was compared under two distraction conditions, electronic reading and voice chat, participants in voice chat were found to perform better in the take-over requests than those in the electronic reading state in a four-second handover condition. Additionally, the participants in the voice chat condition had a more favorable body posture, making it easier to regain manual driving \cite{you2018take}.

\section{Ethics, AV Acceptance and User Perception}
The advent of AVs represents a significant milestone for the automotive industry; however, it is imperative to consider the ethical implications of this technology, particularly for older or disabled individuals. AVs will eventually make decisions without the input of a human driver, making it essential to establish ethical guidelines that prioritize the safety and well-being of all occupants and those in the surrounding environment. Special attention must be given to individuals with special needs who may require additional accommodations or support to ensure their safety. Moreover, concerns about AVs extend beyond ethical considerations, including in-vehicle system malfunctions, knowledge and learning, and functional and hedonic motivation\cite{stiegemeier2022really}. Factors such as education level, age, and attitude toward AV also affect people's perception of AVs. Research suggests that younger individuals with less education tend to be more open to AV and willing to share roads with them. In contrast, older individuals with even higher education appear to be more reluctant\cite{wang2020intelligent}. This may explain why older adults hesitate to use AVs, even if they can quickly adapt to the human-machine interface (HMI).

Studies examining trust and acceptance of highly automated driving systems (HADs) among younger and older drivers have shown that both groups consider HAD trustworthy and acceptable, with trust and acceptance showing comparable developmental patterns over different stages of system experience \cite{hartwich2019first}. Positive initial experiences with HADs were crucial in establishing drivers' trust, acceptance, and system usage. However, the potential risk of over-reliance and misuse of HADs has also been identified. Future research should focus on designing HMIs for HADs that support adequate system experiences from the initial phase to appropriate ongoing usage, particularly for older drivers. Finally, to comprehensively assess the relationship between AVs and ethics, it is essential to consider several key factors, such as criminalization, paternalism, privacy, justice, responsibility, transparency, justice and fairness, non-maleficence, responsibility, privacy, beneficence, freedom, autonomy, trust, sustainability, dignity, and solidarity\cite{ito2022consideration}. Addressing these issues will ensure AVs' safe and ethical operation.

\subsubsection{AV Acceptance}
For elderly and special needs drivers, research and design that considers their limitations must be emphasized. Incorporating advanced assistance systems and improving interior design aspects can help improve usability and safe driving \cite{fernandes2020challenges}. Trust is found to have a positive correlation to perceived safety and driving comfort \cite{hartwich2020passenger}. Because of low trust, automated vehicle control felt less pleasant than human vehicle control in all aspects of the driving experience, thus having a lower system acceptance. It is imperative to provide enough information while driving. Similarly, the term \textit{cutting-edge} had a positive relationship with user acceptance, while bothersome and apprehensive were some emotions having a negative relationship \cite{lee2022effect}. Studies suggest there is a need for marketing future AV taxi services, including providing early service experiences, maximizing differentiation between AV taxi and conventional taxi services, addressing low-reliability issues, and optimizing speed service for individual users. Recent studies have also shown that older drivers experienced better control and driving efficacy while riding in an AV as compared to interacting with a simulator \cite{classen2020older}, indicating that exposure to the AV may result in a superior mode of automation that influences user acceptance by revealing what it's actually like. However, in cases where actual physical experiences are not possible, AV simulators have also been shown to provide initial exposure that increases trust and acceptance\cite{classen2021older, rovira2019looking}. The mobility level in older adults(also majorly affected by cognitive performance) has been shown to significantly and positively affect AV perception and desire for knowledge\cite{park2024automated}. Even higher education individuals tend to have more negative opinions regarding AV safety compared to others\cite{li2024effect}; however, their opinions are more likely to shift towards the positive side after a successful test ride\cite{sharma2022quantifying}. Thus it's imperative to provide any forms of demonstrations and hands-on training for increased trust and perceived safety. With more recent AI developments, AI coaching has been an emerging strategy for training and generating AV experiences. However, one should be careful about the type of information to deliver \cite{kaufman2024effects} for efficient learning. For instance, too much information has been shown to create a sense of overwhelm and more visual information was preferred. However, properly tuned AI coaching has great potential for increased trust, confidence, and expertise.

In the case of BVI individuals, hope for independence and freedom, skepticism about the needs being met, safety concerns, and affordability were identified as major factors that affected the attitude and willingness to use AV \cite{bennett2020willingness}. When developing these technologies, these factors must be considered in the design process from the outset. Moreover, manufacturers also need to consider special policies that let these user groups experience these technologies before mass deployment to confirm whether the accessibility needs are met. Additionally, the definition of accessibility should not only apply to the technological aspect but also the reduction of profit margins the manufacturers make considering the high cost these disabled user groups with relatively low income \cite{blindIncome} need to pay. Thus, public information campaigns must also emphasize the freedom to travel and reassurances concerning safety while addressing affordability concerns. Blind and visually impaired individuals were found to favor the concept of self-driving vehicles and are optimistic about the potential benefits for mobility and independence \cite{brinkley2020exploring}. However, older adults and those with higher levels of education were found to express concerns about their ability to operate the technology and whether their needs were adequately considered in its design. Additionally, visually impaired respondents expressed concerns about legal liability and spurious claims. 

\subsubsection{Trust}
Age has shown to be positively correlated with an individual's initial opinion on AV Safety\cite{li2024effect}. Indeed, people older than 60 are shown to be significantly more concerned about safety \cite{lee2022public, huff2019too} even though they tend to show interest in learning and trying out these technologies \cite{park2024automated}. This greatly affects their willingness to adopt AV technologies in more publicly deployed forms, such as connected and automated vehicles (CAV)
 
\cite{havlivckova2019role}. Using anthropomorphic agents to create a two-way conversational interaction with the user has been found to increase the user's perceived trust and pleasure, with passengers feeling more in control of the journey experience when accompanied by the agent \cite{large2019please}. Using anthropomorphism in the agent's design creates a more 'forgiving' experience, in which passengers are more willing to accept reliability and dependability issues. Modern AI technologies such as GANs have been proven to be effective in generating these types of visual representations\cite{abdal20233davatargan, nagano2018pagan, yi2020audiodriven, boroujeni2024ic, chen2023dh} that can be tailored to a specific user. 

\subsection{Personally Owned and Shared Use vehicle context}
Various factors influence public acceptance of full driving automation for personal and shared-use vehicles. Some prominent factors were safety, compatibility, trust, ease of use, and usage cost. Perceived usefulness, trust, and compatibility were found to have a more significant impact on the behavioral intention to use personally owned concepts than shared-use concepts \cite{motamedi2020acceptance}. 

\begin{table*}[htbp]
    \centering
    \caption{ADAS Technologies with Inclusion and Assistive Classification by Function}
    \resizebox{\textwidth}{!}{%
    \begin{tabular}{l|l|l|l}
        \toprule
         \textbf{Class} & \textbf{Technology (Name)} & \textbf{Control Unit} & \textbf{Inclusion/Assistive} \\
        \midrule
         {\textbf{Safety}}& Automatic Emergency Braking (AEB) & Braking & Assistive \\
         & Collision Avoidance System (CAS) & Steering, Braking & Assistive \\
         & Electronic Stability Control (ESC) & Powertrain, Braking & Assistive \\
         & Pedestrian Detection & Visual Display, Speaker & Inclusion \\
         & Anti-lock Braking System (ABS) & Braking & Assistive \\
         \midrule
         {\textbf{Lane and Speed Assistance}}& Adaptive Cruise Control (ACC) & Powertrain, Braking & Assistive \\
         & Lane Departure Warning (LDW) & Speaker, Visual Display & Assistive \\
         & Lane Keeping Assist (LKA) & Steering & Assistive \\
         & Traffic Sign Recognition (TSR) & Visual Display & Inclusion \\
         \midrule
         {\textbf{Parking Assistance}}& Parking Assistance Systems & Steering, Visual Display & Assistive \\
         & Rear Cross Traffic Alert (RCTA) & Visual Display, Speaker & Assistive \\
         & Blind Spot Monitoring (BSM) & Visual Display, Speaker & Assistive \\
         \midrule
         {\textbf{Driver Monitoring}}& Driver Monitoring System (DMS) & Speaker, Visual Display & Inclusion \\
         & Tire Pressure Monitoring System (TPMS) & Visual Display & Assistive \\
         & Heads-up Display (HUD)* & Visual Display & Inclusion \\
        \bottomrule
    \end{tabular}%
}
    \label{tb:adas_technologies_by_class}
\end{table*}

\section{Inclusion Technology and Policy Frameworks}
This section delineates the relationship between driving inclusion and classical Advanced Driver Assistance Systems (ADAS) to establish a systematic framework for inclusive driving design. While ADAS technologies have significantly improved road safety, driving convenience, and automation, their role in promoting inclusion remains under-explored \cite{chen2024enhancing}. One goal should be to comprehensively evaluate these technologies, analyzing their capabilities, limitations, and potential to support a diverse range of users, including those with disabilities and older adults. To achieve this, focus on existing ADAS technologies such as Adaptive Cruise Control (ACC), Lane Keeping Assist (LKA), and Blind Spot Monitoring (BSM) is required. Each system should be evaluated regarding inclusion and accessibility to highlight its strengths and gaps. By distinguishing technologies that directly enhance inclusion from those primarily focused on general safety and convenience, we aim to provide actionable insights for designing future inclusive driving systems.
This analysis seeks to answer key questions: What defines inclusivity in driving? How do ADAS technologies address or fall short of addressing the needs of vulnerable populations, and what design principles can bridge the gap between general assistance and true inclusion? Through this review, we hope to inspire the development of systematic, user-centric solutions that advance the goals of driving inclusion.

Advanced Driver Assistance Systems (ADAS) encompass a range of technologies designed to enhance safety, convenience, and inclusion in driving. For safety, features like \textbf{Automatic Emergency Braking (AEB)} prevent or mitigate collisions by automatically applying brakes, while \textbf{Collision Avoidance Systems (CAS)} use steering or braking to avoid potential crashes. \textbf{Electronic Stability Control (ESC)} maintains vehicle stability during sharp turns or slippery conditions, and \textbf{Anti-lock Braking Systems (ABS)} prevent wheel lock-up during sudden stops, ensuring better control. Inclusive safety features like \textbf{Pedestrian Detection} identify and react to pedestrians in the vehicle’s path, protecting vulnerable road users. For lane and speed assistance, \textbf{Adaptive Cruise Control (ACC)} dynamically adjusts speed to maintain safe distances, \textbf{Lane Departure Warning (LDW)} alerts drivers if they drift out of their lane, and \textbf{Lane Keeping Assist (LKA)} provides steering inputs to keep vehicles centered. \textbf{Traffic Sign Recognition (TSR)} enhances inclusion by detecting and displaying road signs, ensuring critical information is accessible. Parking assistance technologies, such as \textbf{Parking Assistance Systems}, help drivers maneuver into parking spaces, while \textbf{Rear Cross Traffic Alert (RCTA)} and \textbf{Blind Spot Monitoring (BSM)} reduce collision risks by warning of approaching vehicles or those in blind spots. Driver monitoring systems also play a critical role; the \textbf{Driver Monitoring System (DMS)} tracks driver attention and alertness, and \textbf{Tire Pressure Monitoring Systems (TPMS)} ensures proper tire maintenance by providing real-time alerts. Lastly, HUDs project essential driving information onto the windshield, improving focus and accessibility for drivers by reducing distractions. These technologies create a robust ecosystem for safer and more inclusive driving.

\subsection{Inclusion Categorization}
Despite decades of research and deployment, the traffic and transportation sector has not fully leveraged these advancements to create universally inclusive solutions. This limitation underscores the need for a clear evaluation standard to distinguish assistive technologies from truly inclusive ones. Assistive technologies primarily aim to support specific tasks or user groups, often lacking the flexibility and adaptability to meet the diverse needs of all drivers, particularly those with disabilities. In contrast, inclusive technologies are designed with universal usability, ensuring safety, accessibility, and ease of use across a wide range of users and scenarios.
To address this gap, a systematic framework is essential for evaluating and categorizing technologies based on their inclusion capabilities. Such a framework would provide a robust metric to assess how well current and emerging technologies align with the goals of driving inclusion. Establishing these standards can guide future developments toward solutions that enhance safety and convenience and ensure accessibility and equity in modern transportation systems.

As discussed earlier, the level of inclusion in driving technology often parallels advancements in assistive technology. However, assistive technologies do not always translate into universally inclusive solutions. For instance, Adaptive Cruise Control (ACC) and radar-based dynamic cruising systems offer convenience for individuals with impairments by automating certain driving tasks. Yet, these systems often fall short in emergency response scenarios and, in some cases, may malfunction, leading to serious accidents due to their direct interaction with the drivetrain. This highlights a critical gap in the ability of assistive technologies to address the broader spectrum of user needs.

\subsection{Evaluation of Existing Solutions}
Assistive technologies often focus on individual vehicle safety but may introduce broader inefficiencies. For instance, dynamic cruise control can lead to frequent acceleration and braking, disrupting traffic flow and causing congestion. This proposal emphasizes the importance of developing standardized evaluation metrics to assess the holistic impact of these technologies. Such metrics would identify technologies that genuinely enhance safety and efficiency, distinguish those with neutral effects, and flag those that may introduce systemic harm. Moreover, examining situations where individual safety gains conflict with broader transportation efficiency is imperative, ensuring balanced advancements. 

\subsection{Inclusion Design Vision}
\begin{figure*}[hbt!]
  \centering
\usetikzlibrary{positioning,calc} 
\usetikzlibrary{decorations.text}
\tikzset{shifted by/.style={to path={($(\tikztostart)!#1!90:(\tikztotarget)$)
 -- ($(\tikztotarget)!#1!-90:(\tikztostart)$)}}, 
 shifted by/.default=2pt,standard edge/.style={very thick,-latex}, 
 back and forth between/.style args={#1 and #2}{insert path={
  #1 edge[standard edge,-latex,shifted by] #2 #2 edge[standard edge,shifted by] #1}}} 
\begin{tikzpicture}[font=\rmfamily,
boot/.style={draw,thick,align=center,label=below:#1}, paths/.style={->, thick, >=stealth'}]

 \node[boot=Ingress] (EAB) {Seats Adjust for \\ Space Maximization,\\ Support Handles \\Pop Up};
 \node (Phone) [minimum width=2cm, above= 1cm of EAB] {\includegraphics[height=1cm]{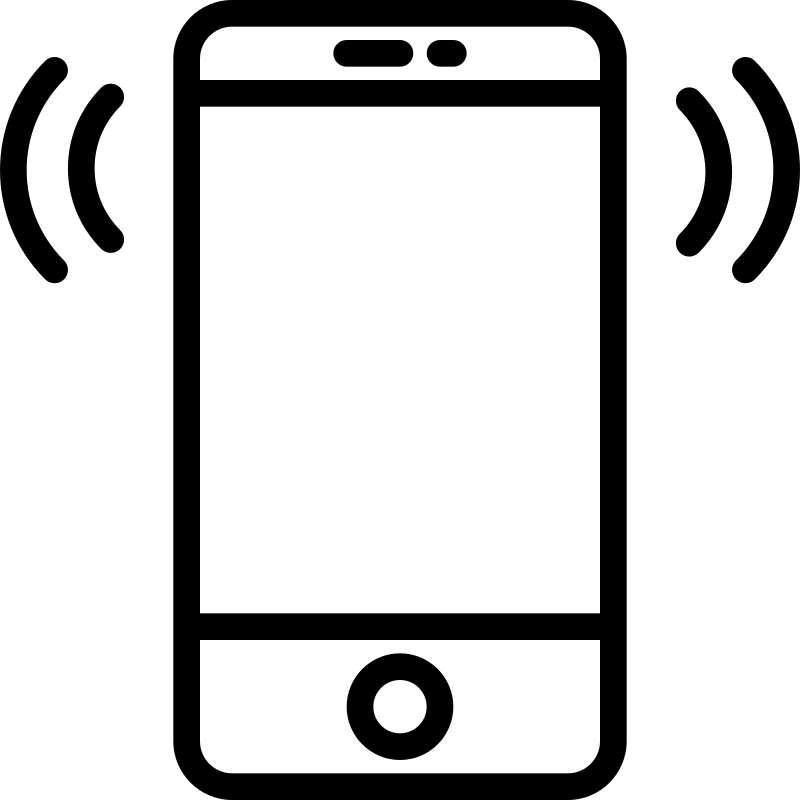}
};
 \node[below=1cm and 0cm of EAB.south,boot=Adaptation/Personalization] (FM) {-Speed\\ -level of kindness\\ -verbosity of system interaction\\ -Region based \\ -Driving Styles \\ -Basic, Technical and \\Supplementary Information \\ - pre-takeover alarm delay \\ -Additional re-engagement assistance};
 \node[right=2cm of EAB, boot=In-vehicle Interaction] (IAB) {UI Adaptation \\ Haptics, Audio, Touch, Visuals};
 \node[below=3cm of IAB.270, boot=Mode based information, align=center] (MBI) {- Timely information on \\current driving mode \\ and vehicle status \\ - status and reason before and after \\ NDRA\\ - Time remaining for NDRA};
 \node[above=1cm and 0cm of IAB.north, boot] (FM2) {Situational Awareness \\ Anthromorphic agents \\NDRA information\\ Current Driving Mode \\ Takeover Warning};
 \node[right=0.2cm and 2cm of IAB.east,boot] (CAB) {Fine \\tune drop off location\\ to avoid extra walking};
 \node[above= 0.1cm of Phone.north, align=center] (profile) {Set user profile\\ using phone};
 \node[below right =1.5cm and 0cm of IAB.east,boot=Additional HMI] (SB) {- Within Driver seat \\or easily accessible \\location \\-HUD interfaces};
 \draw[back and forth between={(IAB.south west) and (FM.north east)},
 back and forth between={(IAB.north) and (FM2.south)},
 back and forth between={([yshift=-0.5cm]IAB.south) and (MBI.north)},
 back and forth between={(Phone.south) and (EAB.north)},
 ];
 \def\myshift#1{\raisebox{-2.5ex}}
 \draw [->,thick,postaction={decorate,decoration={text along path,text align=center,text={|\rmfamily\myshift|Set precise location}}}] (SB.east) to [bend right=35]  (CAB.330);
 
 \def\myshiftR#1{\raisebox{+2.5ex}}
 \draw [->,thick,postaction={decorate,decoration={text along path,text align=center,text={|\rmfamily\myshiftR|Set precise location}}}] (Phone.north east) to [bend left=70]  (CAB.north);

 \draw (EAB) edge[standard edge] (IAB)
 (IAB) edge[standard edge] (CAB) 
 (IAB.south east) edge[standard edge] (SB) 
 ;
 
 \node[above=0.1cm of FM2.north,align=center] (typeDetailComment) {Type/Detail of \\ Presented Information};
 \node[above right=0.1cm of CAB.north,align=center] (egressComment) {Egress};
 \node[below left=0cm of Phone.south,align=center] (ingressComment) {Fine tune\\ pickup location};
 \draw[<->, line width=0.4mm] (Phone)--(IAB.north west) node[midway,sloped, above] {Interact with the IVS} node[midway,sloped, below] {Sync User Profile};; 
\end{tikzpicture}

  \caption{Proposed Framework for end-to-end in-vehicle interaction demonstrates how to design inclusive in-vehicle interaction experience in the proposed paper.}
\label{tab:figureProposedFramework}
\end{figure*}

A proposed framework in Table \ref{tab:ProposedFramework} provides a detailed overview of the design considerations that support various interaction phases, from entering the vehicle to exiting and every interaction. We also present various factors that need to be considered during this phase, such as providing smartphone-based controls for fine-tuning ingress and egress, personalizing the HMI interface based on the user's needs, and providing situational awareness of the surroundings. Additionally, we recommend incorporating multi-modal interactions, using anthropomorphic agents to improve conversational interaction, and providing region-based adaptation for the HMI. Furthermore, the framework highlights the importance of designing takeover warnings using multi-modal feedback, adjusting pre-takeover alarm periods based on the age group, and providing additional assistance for quick driver re-engagement. The table also emphasizes the need to provide enough mode information to avoid state anxiety, allow quick engagement, and articulate driving mode while considering NDRA impact to adjust takeover warning delay. Overall, the table provides a comprehensive list of considerations that should be taken into account by designers and developers of AVs to ensure a safe, comfortable, and personalized driving experience for all users. The information presented in the table is valuable for researchers, practitioners, and policymakers working in AVs. It provides an essential guide to designing effective and efficient AVs that meet the needs of all users. We also visualize and present this framework in Fig \ref{tab:figureProposedFramework}.

We also propose a comprehensive framework, outlined in Table \ref{tab:AdditionalConsideration}, that encompasses key considerations for enhancing the acceptance of AVs. This framework highlights four major components crucial to addressing the barriers related to AV usage among various groups. The first component focuses on identifying user needs to ensure that AVs are designed to cater to the specific requirements of diverse user groups. The second component involves developing strategies to train individuals to become familiar with AV technology, particularly those who may be less inclined or able to adapt to new technological advancements. The third component emphasizes the importance of trust and affordability, which are significant barriers to the widespread adoption of AVs. Addressing safety, privacy, and security concerns is essential to increase trust.
Furthermore, affordability is crucial in making AVs accessible to everyone, regardless of socioeconomic status. Finally, the fourth component highlights the need for design standardization at the government level\cite{razi2023deep}. By establishing a consistent in-vehicle interaction standard across various vendors, users can learn once and easily use AV services from different providers. Overall, our proposed framework serves as a roadmap for stakeholders to enhance the acceptance of AVs among diverse user groups. We believe that addressing the key considerations outlined in our framework can pave the way for a future where AVs significantly improve transportation efficiency and safety.

\section{Missing Gaps and Open Problems}
There is a notable delay in integrating advanced accessibility features, already implemented in devices like a regular smartphone, into a vehicle interaction system.  One reason might be the cost of ownership, as only higher-end vehicles incorporate some of these features while the affordable ones keep them minimal. Even with that, the accessibility features in a regular smartphone are much more capable than those found in expensive vehicle models. This disparity highlights a significant gap in policies that should mandate such accessibility features, at least the basic ones. However, with advancements in AI, we can expect future vehicles to approach a high level of autonomy and interaction to become more adaptive and responsive to user needs through software enhancements instead of hardware limits as in the status quo.

However, significant gaps remain that need attention, even though AI is advancing faster than any technologies we've ever seen. One key issue is the performance inconsistency. While many tools perform well in most scenarios, they may fail in certain others, and even a few failures can substantially impact user interaction. The explainability of actions or responses generated is another challenge currently being addressed. Sometimes, these tools may generate unpredictable outputs without logical reasoning. It has also been evident that users often overtrust AI, which might lead to significant setbacks if it fails.

The pace of AI development is outpacing regulatory frameworks, creating uncertainty about accountability in cases of system failures. Legal and ethical questions about liability, particularly in autonomous systems, remain unresolved, posing challenges for widespread adoption in accessibility-focused applications. Interoperability is yet another issue that needs to be addressed. There is a lack of standards for how different AI and accessibility systems should interact both within the vehicle and with external environments like pedestrians and other vehicles. This can result in disjointed user experiences, particularly when moving between different vehicle manufacturer systems. 

Robust real-world testing is also another major issue. Many AI tools and systems are evaluated under ideal conditions that fail to reflect the complexity and variability of real-world scenarios. For instance, it is essential for AI systems in vehicle perception to have inherent robustness to unforeseen circumstances like varying weather conditions\cite{bastola2024robustformer}. This gap in testing and validation might lead to unexpected failures when deployed at scale, especially in critical environments like AVs.

Lastly, the accessibility gap is further widened by the digital divide. Users from low-income or rural areas often lack access to the infrastructure required to support these advanced systems, such as reliable internet connectivity or compatible hardware. Without addressing these fundamental barriers, even the most advanced AI solutions risk becoming inaccessible to large sections of the population. Bridging these gaps requires a multi-pronged approach, encompassing policy reform, inclusive design practices, thorough testing protocols, and efforts to ensure equitable access. Collaborative initiatives between industries, governments, and advocacy groups are also crucial to unlocking the full potential of AI-driven accessibility technologies.

These challenges highlight the need for further development and oversight of AI technologies used in accessibility contexts, such as AVs, to ensure they are reliable and transparent. 
    
\section{Discussion}
We presented a comprehensive review of driving inclusion, distinguishing between assistive technologies and genuinely inclusive solutions. However, a few more terms need to be discussed as these topics have not been well covered in other papers. While they mostly focus on the technical aspect, but are very important to not only the human-centered computing and interaction areas but also the autonomous driving and transportation. While this review provides insights into driving inclusion, several limitations must be acknowledged. The majority of research focuses on older adults and visually impaired individuals, with limited exploration of other disabilities. Furthermore, the variability in cognitive and physical abilities among users presents challenges in designing universally inclusive systems.

\subsection{Relation with ADAS}
This study comprehensively reviews driving inclusion, distinguishing between assistive technologies and genuinely inclusive solutions. While assistive technologies, such as Adaptive Cruise Control (ACC) and Automatic Emergency Braking (AEB), enhance safety and convenience, they often fail to meet the broader needs of diverse user groups, including individuals with disabilities and older adults. For instance, ACC offers dynamic cruising capabilities but may fail during emergencies or cause safety risks due to direct drivetrain interactions.

This distinction underscores the need for a systematic evaluation framework to categorize technologies based on their inclusivity. Truly inclusive technologies prioritize universal usability beyond assistive solutions to accommodate the unique needs of underrepresented populations. This review establishes these distinctions and provides a foundation for developing technologies that advance driving inclusion.

\subsection{Relation with Autonomous Driving}
Vehicle inclusion design is closely tied to the level of autonomy. At lower levels (Levels 1-2), inclusion efforts are primarily focused on safety features such as AEB, Lane Departure Warning (LDW), and Blind Spot Monitoring (BSM). These features enhance safety and reduce driver fatigue but do not fully accommodate users unable to perform driving tasks.

Moderate autonomy (Level 3) introduces capabilities like highway lane-keeping and adaptive speed adjustments, which address inclusion by reducing physical and cognitive demands. However, manual intervention in complex scenarios still limits accessibility. Higher autonomy levels (Levels 4-5) hold the greatest potential for inclusion, offering hands-free operation and features like automated lane merging and full self-driving capabilities. These advancements enable individuals with disabilities and older adults to achieve greater independence. Nonetheless, legal uncertainties around liability for fully autonomous systems pose significant barriers to widespread adoption and inclusive design implementation.

\subsection{Relation with Transportation}
Inclusive design in autonomous driving extends its impact on transportation networks, reshaping traffic systems at a systemic level. Advanced driving technologies contribute to safer and more efficient transportation, even at basic autonomy levels. For instance, Toyota's Dynamic Radar Cruise Control (DRCC) and Lane-Centering Assist enhance road safety by reducing human error. Similarly, BMW’s use of traffic cameras and data-sharing technologies optimizes real-time route planning, alleviating congestion.

Modern vehicles’ Vehicle-to-Everything (V2X) communication capabilities transform cars into active participants in interconnected transportation ecosystems. This interoperability enables dynamic traffic management, continuously monitoring and adjusting road conditions and vehicle locations. As more vehicles adopt autonomous and connected technologies, urban transportation systems can become safer, more efficient, and inclusive, benefiting the entire mobility ecosystem.

\subsection{Ethical Considerations}
The pursuit of inclusive autonomous vehicle design introduces complex ethical and liability challenges. Autonomous systems have the potential to revolutionize mobility by reducing the reliance of older adults and individuals with disabilities on traditional driving skills. However, the widespread adoption of inclusive features is hindered by liability concerns. Determining fault—whether human or machine—remains legally ambiguous when incidents occur, deterring manufacturers from promoting inclusive technologies.

Additionally, automation complacency presents a unique ethical challenge. Unlike system failures, automation complacency arises from human behavior, where drivers over-rely on automated systems and neglect their responsibility to monitor and intervene. For instance, a driver assuming that AEB or ACC will manage all situations may fail to take necessary actions during emergencies. This over-reliance, particularly at intermediate automation levels (Levels 2-3), risks undermining safety and diminishing user awareness of system limitations.

Addressing these ethical concerns requires regulatory clarity to allocate liability appropriately and foster user education to balance technological capability with human responsibility. Without these measures, the potential for autonomous systems to enhance inclusion and equity will remain underutilized.

\section{Conclusion}
This paper has presented a comprehensive discussion of designing an inclusive in-vehicle human-machine interface, focusing on addressing the needs and abilities of older adults and individuals with various disabilities. Our proposed end-to-end framework for in-vehicle interaction incorporates various technologies to facilitate the actions a driver might need to take to improve user acceptance of AVs. Our findings have important implications for researchers and manufacturers designing inclusive in-vehicle interaction systems. Specifically, our framework provides a valuable reference point for developing more accessible and personalized interfaces that accommodate all individuals' diverse needs and abilities. Additionally, our work provides insight for researchers studying user needs and ways to improve the acceptance of AVs.

The proposed architectural design offers guidelines for developing more inclusive in-vehicle interaction systems to enhance the overall driving experience for all individuals. Noting that many assistive technologies are still required to cover the diverse needs of passengers with a disability, we hope our review of the current practice and lacking features will inspire further research in this area and lead to more innovative solutions to improve the lives of older adults and individuals with disabilities.

\bibliographystyle{plain}
\bibliography{references}  
\end{document}